\documentclass[preprintnumbers,
amsmath,amssymb,floatfix,10pt,prd,twocolumn,
superscriptaddress,longbibliography,nofootinbib]{revtex4-2}
\usepackage{bm}
\usepackage{amsfonts}
\usepackage{latexsym}
\usepackage{amsmath}

\usepackage{palatino}
\usepackage{mathpazo}
\usepackage{textcomp}
\usepackage{float}
\usepackage{booktabs}
\usepackage{dcolumn}
\usepackage{ragged2e}
\usepackage{epsfig}
\usepackage[dvipsnames]{xcolor}
\usepackage{hyperref}
\hypersetup{           
    colorlinks=true,                
    breaklinks=true,                
    urlcolor= OrangeRed,                
    linkcolor= magenta,                
    bookmarksopen=false,
    filecolor=black,
    citecolor=blue,
    linkbordercolor=blue}
\usepackage{graphicx}
\usepackage{orcidlink}
\usepackage[T1]{fontenc} 

\begin{document}
\title{Shadows of a Non-Commutative Black Hole under the Influence of a Magnetized Plasma}
\author{Mrinnoy M. Gohain\orcidlink{0000-0002-1097-2124}}
\email{mrinmoygohain19@gmail.com}
\affiliation{%
 Department of Physics, Dibrugarh University, Dibrugarh \\
 Assam, India, 786004}
\affiliation{%
 Department of Physics, DHSK College, Dibrugarh \\
 Assam, India, 786001}
 
\author{Kalyan Bhuyan\orcidlink{0000-0002-8896-7691}}%
 \email{kalyanbhuyan@dibru.ac.in}
\affiliation{%
 Department of Physics, Dibrugarh University, Dibrugarh \\
 Assam, India, 786004}%
 \affiliation{Theoretical Physics Divison, Centre for Atmospheric Studies, Dibrugarh University, Dibrugarh, Assam, India 786004}

%
\author{Paragjyoti Chutia \orcidlink{0009-0005-3897-0293}}%
\email{paragjyotic@gmail.com}
\affiliation{%
 Department of Physics, DHSK College, Dibrugarh \\
 Assam, India, 786001}
 
%
%
%

\begin{abstract}
We investigate the properties of the shadow cast by a black hole (BH) in a non-commutative spacetime surrounded by a plasma medium. BHs in a plasma environment can exhibit complex interactions that profoundly affect their shadow properties. In specific terms, we consider a special type of plasma medium motivated from the interactions among axions, photons and plasma excitations, or commonly termed as axion-plasmon in the presence of a magnetic field. We have studied the effect of these plasma parameters and the non-commutative parameter on the shadow properties by focusing particularly on the photon sphere radius, angular shadow and the shadow observed from the observer's perspective. For this we have considered two different forms of plasma, \emph{viz.} homogeneous and inhomogeneous plasma. In this work we have presented the parameter dependence of these BH shadow properties. Finally, we have also studied the energy emission rate of the BH and how it depends on the model and plasma parameters.
\end{abstract}

\maketitle
\section{Introduction}
BHs are regions of extreme curvature in spacetime predicted by general relativity (GR), are one of the most fascinating entities in the universe. A key motivation for studying BHs is to test GR in the strong-field regime and to understand the fate of collapsing massive stars. Furthermore, BHs play an important role in the evolution of galaxies, often by residing at the galactic centers and by influencing the dynamics of surrounding stars and gas.

Recent works on BHs explore various aspects, including the resolution of singular points of infinite density at the center of classical BHs. One of such approaches deals with non-commutative geometry of spacetime, which introduces a fundamental "fuzziness" to spacetime at extremely small scales \cite{Nicolini2009Mar,Schneider2020Jul,Bastos2010Apr}. This is termed loosely as "smearing of mass". 
The smearing of mass, inspired by non-commutative geometry, helps avoid singularities in BH solutions by preventing the gravitational collapse of matter into an infinitely dense point. Instead of a point-like singularity, the mass distribution becomes spread out over a finite region, often modelled by a Gaussian or Lorentzian distribution \cite{Nicolini2009Mar}.  Non-commutative geometry introduces an intrinsic uncertainty in the localization of points in space-time, which fundamentally  suggests how matter and energy may be distributed, especially in situations with strong gravitational fields. As a result, singularities, which arise from assuming point-like matter distributions, may be avoided by considering non-local matter distributions. This concept of smearing of the mass distribution plays a significant role in the curvature of spacetime in the manner that, classically, in a BH, the curvature becomes infinite at the singularity but with a smeared mass distribution, the spacetime curvature stays finite even at the centre of the BH. This produces a regular BH solution. Speaking of regular BHs, In the recent years, several new families of regular BH solutions have been constructed and analyzed in detail. Ovalle et al. \cite{OvalleAraya2023} introduced regular hairy BHs via a Minkowski-deformation procedure, yielding non-singular extensions of Schwarzschild and Kerr spacetimes that satisfy energy conditions and possess well-defined horizons without central singularities. Pedrotti and Vagnozzi \cite{Pedrotti2024} then extended the eikonal quasinormal mode shadow correspondence to rotating regular BHs, deriving exact analytical relations for axisymmetric spacetimes and verifying them explicitly for the rotating Bardeen and Hayward metrics. Junior et al. \cite{Junior2024} constructed regular solutions in conformal Killing gravity coupled to nonlinear electrodynamics and scalar fields, showing that the Kretschmann scalar remains finite everywhere and interpolates between a de Sitter core and an asymptotically flat exterior. Zhang et al. \cite{Zhang2025} built fully regular Bardeen–Dirac stars in AdS spacetime with static, spherically symmetric configurations supported by both electromagnetic and Dirac fields and explored their ADM mass, Noether charge, and light-ring structure.
Additionally, Capozziello et al. \cite{Capozziello2024,DeBianchi2025Jun} showed that Lorentzian–Euclidean BH metrics avoid singularities via a signature-change mechanism (``atemporality'') which maintain finite curvature invariants throughout the manifold . Many other works related to regular BHs can be found in Refs. (and references therein).~\cite{Nicolini2017Sep,Vertogradov2025Jun,Vertogradov2025May,Calza2025Jul,Abbas2024Apr,Bueno2025May,
Huang2025May,Bueno2025Feb,Frolov2025Feb,Bueno2025May1}.

Quantitatively,
the extent of the smeared region is parametrised by the non-commutativity parameter, usually represented as $\sqrt{\theta}$, which introduces an intrinsic length scale below which the classical definition of a point like object becomes redundant.  The basis for the application of non-commutative geometry to BHs is the idea that spacetime coordinates cannot be measured with infinite accuracy. The amount of energy needed for such accurate measurements would necessarily change the geometry at these scales, introducing an inherent precision in coordinate measurements. This is somewhat analogous to the Heisenberg uncertainty principle in quantum mechanics. Some important works carried out recently using non-commutative spacetime includes the work of Toghrai et al \cite{Toghrai2023Dec} where they studied the non-commutative formalism of a Schwarzschild BH and some of its physical features. Bhar et al \cite{Bhar2025Apr} investigated the thermodynamical properties, shadows, quasinormal modes of the non-commutative BH in the framework of de Rham-Gabadadze-Tolley like massive gravity. Some other works can be found in Refs.~\cite{Afshar2025Feb,S.Afshar2025May,
Wang2025Feb,Quevedo2025Feb,Tan2025May,Filho2025Feb,AraujoFilho2025Feb,BahrozBrzo2025Mar,
Ashraf2025May,Wang2025Mar,Filho2025Feb,Hamil2025Mar}.

The surroundings of a galactic BH can contain matter in the form of plasma. This is because, as matter accretes onto a BH, it is heated by gravity and friction to extremely high temperatures in a hot accretion disk. The heat ionises the accreting matter, thus forming a plasma that emits energy over the entire electromagnetic spectrum, and can be an interesting probe from the observational point of view. The plasma environment can also be a useful ingredient in a wide range of astrophysical processes, including the production of relativistic jets from the BH poles and the production of strong magnetic fields. There have been many works dedicated to the effect of plasma on the optical properties of BHs, for instance in a work by Atamurotov et al, \cite{Atamurotov2015Oct}, the authors studied the motion of photons around a Kerr BH in a plasma with a radial power-law parametrisation of density and found that the BH shadow size and shape are dependent on the plasma parameter, spin, and inclination angle, while in a Schwarzschild BH, the photon sphere remains unchanged, but the shadow size is reduced by refraction, and the highest energy emission rate gets reduced in the presence of plasma. In another work by Perlick et al, \cite{Perlick2015Nov}, they studied the effect of a non-magnetized plasma on the shadows in spherically symmetric spacetimes, and they derived general formulas which they used in studying Schwarzschild BHs and Ellis wormholes, including free-fall plasma cases and discussed the observational perspectives on the shadows of supermassive BHs. Abdujabbarov et al \cite{Abdujabbarov2015Dec} discussed the shadow of a rotating BH with quintessential energy in both vacuum and plasma and found that in vacuum, the quintessential field enlarges the radius of the shadow and diminishes its distortion, whereas in a plasma medium, the shape of the shadow depends crucially on the plasma parameters, spin, and the quintessential field. There are several other works dealing with the effect of plasma on BH properties and can be found in \cite{Huang2018Jun,Chowdhuri2021Sep,Wang2021Jun,
Badia2022Aug} etc.

Furthermore, plasma around a BH may also couple to theoretical particles like axions (which were proposed to solve the strong CP problem in particle physics) through superradiance, draining energy and angular momentum from spin of BHs \cite{Arvanitaki2011Feb,Arvanitaki2015Apr}. Superradiating axions would produce an axion cloud dense near the BH. Plasma can have a significant role in this phenomenon as well as in the development of the axion cloud. Although not a standard terminology, the "axion-plasmon" terminology reflects the interesting interaction among axions, photons and plasma excitations near the BH environment. Plasma influences, like screening electric charges and modifications of the photon propagation \cite{Balakin2013Dec}, can affect the interaction between axions and the electromagnetic field and may produce observable signatures \cite{Lyutikov2021Aug,Sen2018Nov,Tercas2018May}. Analysis of these couplings has the potential to add valuable perspectives to both axion physics as well as plasmas in a strong gravitational field. Some of the important works based on the effect of axion-plasmon on the optical properties of BHs include the works of Atamurotov et al \cite{Atamurotov2021Sep}, where they studied the effect of axion-plasmon coupling on the null trajectories around a Schwarzschild BH and how it affects the BH shadow, deflection angle and Einstein rings caused by radially infalling gas on the BH in the presence of a plasma medium. They found that these optical properties are influenced by the axion-plasmon coupling parameter. They also showed that the size of the BH shadow decreases with the increase in the axion-plasmon parameter if the observer's position lies sufficiently far away. Khodadi \cite{Khodadi2022Dec} investigated the shadow of a spinning BH within an axion-plasmon cloud and observed that axion-photon interaction affects the shadow size and shape. He found that with increasing value of axion-plasmon coupling, the size of the shadow increases for high-spinning BHs, whereas more massive axions show smaller impacts; in the non-rotating case, the shadow size diminishes, and the axion-plasmon cloud boosts energy emission for rotating BHs but seen to decrease as the rotation slows down.
Pahlavon \cite{Pahlavon2024Jul} studied the effect of a plasma medium on a Riessner-N\"{o}rdstr\"{o}m BH where the effects of axion-plasmon coupling on the optical features of a charged BH is investigated, and they also studied the photon geodesics, shadow radius, deflection angle, Einstein rings. It was shown that the shadow size decreases as charge increases, whereas its appearance varies with the plasma model in consideration, and it is maximum without plasma and minimum in the case of homogeneous plasma. Some other progress in the arena of BHs in a plasma environment over the recent years includes the work of Ali et al \cite{Ali2025Apr} where they studied deflection of light and shadow properties of a hairy BH under the effect of non-magnetic plasma. Rukkiyya and Sini \cite{VP2025Mar} investigated strong lensing properties and shadows of a quantum Schwarzschild BH in the presence of a homogeneous plasma medium. Umarov et al \cite{Umarov2025May} studied the optical properties under the influence of plasma medium around a Sen BH, particularly on the shadow, weak lensing, and time delay properties. Alimova et al \cite{Alimova2025Feb} investigated weak lensing and shadows in the framework of Kalb–Ramond gravity in Reissner–N\"{o}rdstr\"{o}m spacetime using uniform plasma, singular and non-singular isothermal spheres. Some other relevant works can be found in \cite{Ibrokhimov2025Feb,Turakhonov2025May,
Mushtaq2025Jan,Alloqulov2025Apr,Yasmeen2025Feb,Ali2025Mar}.

The paper is organized as follows: In Section \ref{Sec2} we briefly review the physics of the axion-plasmon in a spherically symmetric non-commutative spacetime. In Section \ref{sec3}, we calculate the photon sphere radii for different types of plasma medium. In Section \ref{sec4}, we study the shadow properties by calculating the angular and shadows in a cartesian plane and studied them numerically. In Section \ref{sec5}, obtained observational bounds on the model parameters with reference to the Sgr* shadow data. In Section \ref{sec6}, we investigated the energy emission rate of the non-commutative BH in the presence of plasma medium. Finally, in Section \ref{sec7}, we summarize our study with the conclusion.

\section{Photon motion around the BH in the presence of axion-plasmon}
\label{Sec2}
In this work, we want to explore the consequences of the optical properties of a non-commutative BH in a generalized electromagnetic theory which includes the axion-photon interaction in it \cite{Mendonca2020Mar,Wilczek1987May,Pahlavon2024Jul}. As a result, one needs to introduce an extra term, that accounts for the axion-photon interaction. The Einstein-Maxwell action is given by 
\begin{equation}
    S=\int \left(\frac{R}{16 \pi}- \frac{1}{4}F_{\mu\nu}F^{\mu\nu}\right) \sqrt{-\mathtt{g}}\, d^4x+S_{\rm matter},
    \label{action}
\end{equation}
where 
\begin{eqnarray}
    S_{\rm matter}=\int \mathcal{L}_{\rm matter}\sqrt{-\mathtt{g}}\, d^4x,
\end{eqnarray}
with 
\begin{eqnarray}
    \mathcal{L}_{\rm matter}=\mathcal{L}_\varphi-A_\mu J_e^\mu+\mathcal{L}_{\text{int}}.
\end{eqnarray}
Here, $\mathcal{L}_{matter}$ is the total matter Lagrangian that is composed of the axion Lagrangian $\mathcal{L}_{\varphi}$, the interaction lagrangian $\mathcal{L}_{int}$ and the Lorentz scalar term $A_\mu J_e^\mu$. Also, $\mathtt{g}$ denotes the determinant of the metric tensor.

In our work, we shall assume that the surrounding plasma does not significantly impact on the overall curvature of spacetime. We are only interested in, what role the plasma medium plays in determining the photon trajectories. The interaction between the electromagnetic field and the plasma can affect the energy shifts of the photons, which probably can dictate the optical properties around the BH.
In the action given by Eq. \eqref{action} above, \( J_e^\mu \) represents the four-current associated with electrons, while \( F_{\mu\nu} \) is the electromagnetic field tensor, that embeds the electric and magnetic field components. The axion Lagrangian contribution is given by \( \mathcal{L}_\varphi= \nabla_\mu\varphi^*\nabla^\mu\varphi-m_\varphi^2|\varphi|^2 \), which describes the dynamics of the axion field, including the kinetic and mass terms. The interaction term \( \mathcal{L}_{\text{int}}=-(g/4)\varepsilon^{\mu \nu\alpha\beta}F_{\alpha\beta}F_{\mu \nu} \) represents the interaction between axions and photons.  This term, where \( g \) denotes the strength of the axion-photon coupling, describes how axions affect the electromagnetic field through a pseudoscalar interaction~\cite{Pahlavon2024Jul}.

The static spherically symmetric spacetime metric ansatz is 
\begin{equation}\label{metric}
ds^2=-f(r)dt^2+\frac{1}{f(r)}dr^2+r^2(d\theta^2+\sin^2{\theta}d\phi^2),
\end{equation}
where $f(r)$ is the lapse function.

In the context of noncommutative geometry-inspired gravity, the usual point-like mass profile of the Schwarzschild BH is replaced by a smeared mass distribution. This approach effectively incorporates the minimal length scale of non-commutative geometry, regularizing curvature invariants and removing the traditional singularity at the BH core. In our analysis, we adopt the Lorentzian distribution for matter density. 

For a gravitational source characterized by a Lorentzian-smeared mass distribution, the mass density is defined as \cite{Nicolini2009Mar,Schneider2020Jul,Bastos2010Apr,Feng2024Nov}
\begin{equation}
\rho_\theta(r) = \frac{\sqrt{\theta}\, M}{\pi^{3/2}\, (\theta + r^2)^2},
\end{equation}
where $M$ is the total mass of the BH and $\theta$ is the non-commutative parameter representing the scale of nonlocality. In non-commutative BH solutions, $\theta$ determines a minimum horizon radius $r_0 \sim \sqrt{\theta}$. This smears the central mass over that scale, which removes the classical singularity at $r=0$~\cite{Nicolini2009Mar}. 
At the end of Hawking evaporation, it establishes a stable remnant of size $r_{\rm rem} \sim \sqrt{\theta}$ (with mass $M_0 \sim M_P^2 \sqrt{\theta}$), thus curing temperature divergence at small horizon radii. It also defines the energy scale $\Lambda_{\rm NC} \sim 1/\sqrt{\theta}$, where spacetime “fuzziness” becomes important~\cite{Nicolini2009Mar}.

The cumulative (smeared) mass within a radius $r$ is \cite{Feng2024Nov}
\begin{equation}
\begin{aligned}
\mathcal{M}_\theta(r) &= \int_0^r 4\pi r'^2 \rho_\theta(r')\, dr' \\
&= \frac{2M}{\pi} \left[ \arctan\left(\frac{r}{\sqrt{\pi \theta}}\right) - \frac{r \sqrt{\pi \theta}}{\pi \theta + r^2} \right].
\end{aligned}
\end{equation}
In the large-$r$ limit, the smeared mass approaches $M$, while for small $r$ the mass is regular and non-zero.

The Schwarzschild metric’s lapse function under noncommutative geometry (without perfect fluid dark matter) is modified by replacing the point mass with the smeared mass profile
\begin{equation}
f(r) = 1 - \frac{2 \mathcal{M}_\theta(r)}{r},
\end{equation}
i.e.,
\begin{equation}
f(r) = 1 - \frac{4M}{\pi r} \left[ \arctan\left(\frac{r}{\sqrt{\pi\theta}}\right) - 
\frac{r\sqrt{\pi\theta}}{\pi \theta + r^2} \right].
\end{equation}
This formulation ensures the regularity of spacetime at the origin, as the mass function no longer diverges for $r \to 0$.
For small values of $\theta$, the lapse function can be expanded and approximated to ~\cite{Feng2024Nov}
\begin{equation}
f(r) = 1 - \frac{2M}{r} + \frac{8\sqrt{\theta}M}{\sqrt{\pi} r^2},
\end{equation}
which shows the correction to the Schwarzschild solution due to non-commutative effects.

We point out that the explicit expansion (using Taylor's and Binomial expansion) of the smeared mass function shows that it behaves as $\mathcal{M}_\theta(r) \sim r^3$ as $r \to 0$, despite the fact that the expression for the lapse function contains a $1/r$ term. This indicates that there is no singularity at the center. Therefore, because of the noncommutative matter distribution, all geometric and physical quantities are regular everywhere, and the metric function stays finite at the center.

If we are interested in photon trajectories about a BH in the presence of axion-plasmon medium, the Hamiltonian can be defined as  \cite{Synge:1960b}
\begin{equation}
\mathcal H(x^\alpha, p_\alpha)=\frac{1}{2}\left[ g^{\alpha \beta} p_\alpha p_\beta - (n^2-1)( p_\beta u^\beta )^2 \right],
\label{generalHamiltonian}
\end{equation}
Here, the spacetime coordinates, four-velocity and four-momentum of the photon are defined by $x^\alpha$ $u^\beta$ and $p_\alpha$ respectively. Also, $n$ is the refractive index of the plasma medium ($n=\omega/k$, with $k$ as the wave number). The refractive index of the plasma medium in the framework with the axion-plasmon term is given by  \cite{Mendonca2020Mar}
\begin{eqnarray}
n^2&=&1- \frac{\omega_{\text{p}}^2}{\omega^2}-\frac{f_0}{\gamma_{0}}\frac{\omega_{\text{p}}^2}{(\omega-k u_0)^2}-\frac{\Omega^4}{\omega^2(\omega^2-\omega_{\varphi}^2)}\nonumber \\
&&-\frac{f_0}{\gamma_{0}}\frac{\Omega^4}{(\omega-k u_0)^2(\omega^2-\omega_{\varphi}^2)},
\label{eq:n1}
\end{eqnarray}
where 
$$\omega^2_{p}(x^\alpha)=4 \pi e^2 N(x^\alpha)/m_e$$ 

is the plasma frequency in which $e$, $m_e$ and $N$ denotes the electron charge, mass and number density of electrons. Here, $\omega(x^\alpha)$ is the photon frequency defined as $\omega^2=( p_\beta u^\beta )^2$; the axion frequency is denoted by $\omega_{\varphi}^2$
; the axion-plasmon coupling parameter is given by $\Omega=(gB_{0}\omega_{p})^{1/2}$ with $B_0$ as the homogeneous magnetic field directed in the $z$-direction, where \(g\) represents the axion--photon coupling~\cite{Mendonca2020Mar}\footnote{Here, $g$ should not be confused with the determinant of the metric tensor.}. Given the axial symmetry of the magnetic field and the spherically symmetric nature of the spacetime, we shall assume that the effects of strong curvature predominantly govern the particle dynamics, thereby dominating the influence of the magnetic field.  The  fraction of electron beam moving inside the plasma with velocity $u_0 $ is given by $f_0$, where $\gamma_0$ is the associated Lorentz factor. As we are uncertain about the significance of the electron beam scenario in the vicinity of the BH, it will be convenient for us to set $f_0=0$ for the sake of simplicity. Then, we can rewrite (\ref{eq:n1}) as \cite{Atamurotov2015Oct}
\begin{eqnarray}
n^2(r)&=&1- \frac{\omega_{\text{p}}^2(r)}{\omega(r)^2}-\frac{\Omega^4}{\omega(r)^2[\omega(r)^2-\omega_{\varphi}^2]},\nonumber \\
&&=1- \frac{\omega_{\text{p}}^2(r)}{\omega(r)^2}\left(1+\frac{ g^2B^2_0 }{\omega(r)^2-\omega_{\varphi}^2}\right)
,
\label{eq:n2}
\end{eqnarray}
with
\begin{equation}
\omega(r)=\frac{\omega_0}{\sqrt{f(r)}},\qquad  \omega_0=\text{const}.
\end{equation}
For the experimental requirements for axion-plasmon conversion, the plasma frequency scale must satisfy the inequality \(\omega_{\text{p}}^2 \gg \Omega^2 \quad \text{(or equivalently,} \quad \omega_{\text{p}} \gg gB_0\text{)},\).  According to \cite{Perlick2015Nov}, the photon energy at spatial infinity is given by \(\omega(\infty) = \omega_0 = -p_t,\) and the lapse function is defined such that \(f(r) \to 1 \quad \text{as} \quad r \to \infty.\) In addition, for distinguishing the BH shadow from the vacuum case, the plasma frequency must continue to be much smaller than the photon frequency,  i.e. \(\omega_{\text{p}}^2 \ll \omega^2 \).  Now, for light ray propagation in an axion-plasmon medium, the Hamiltonian can be recast in the following way: \cite{Synge:1960b,Rog:2015a}
\begin{equation}
\mathcal{H}=\frac{1}{2}\Big[g^{\alpha\beta}p_{\alpha}p_{\beta}+\omega^2_{p}\Big(1+\frac{ g^2B^2_0 }{\omega^2_0-\omega_{\varphi}^2}\Big)\Big]. \label{eq:hamiltonnon}
\end{equation}
 The four velocity components for the photons in the equatorial plane $(\theta=\pi/2,~p_\theta=0)$ are:
\begin{eqnarray} 
\dot t\equiv\frac{dt}{d\lambda}&=& \frac{ {-p_t}}{f(r)}  , \label{eq:t} \\
\dot r\equiv\frac{dr}{d\lambda}&=&p_rf(r) , \label{eq:r} \\
\dot\phi\equiv\frac{d \phi}{d\lambda}&=& \frac{p_{\phi}}{r^2}, \label{eq:varphi}
\end{eqnarray}
where we have used the formula for obtaining four-velocity using the Hamiltonian, $\dot x^\alpha=\partial \mathcal{H}/\partial p_\alpha$. Using Eqs. (\ref{eq:r}) and (\ref{eq:varphi}), we can form a general equation for null trajectories
\begin{equation}
\frac{dr}{d\phi}=\frac{g^{rr}p_r}{g^{\phi\phi}p_{\phi}}.    \label{trajectory}
\end{equation}
The constraint equation $\mathcal H=0$, sets the above equation as ~\cite{Perlick2015Nov}
\begin{equation}
 \frac{dr}{d\phi}=\sqrt{\frac{g^{rr}}{g^{\phi\phi}}}\sqrt{h^2(r)\frac{\omega^2_0}{p_\phi^2}-1},
\end{equation}
where
\begin{equation}
    h^2(r)\equiv-\frac{g^{tt}}{g^{\phi\phi}}-\frac{\omega^2_p}{g^{\phi\phi}\omega^2_0}\left(1+\frac{ g^2B^2_0 }{\omega^2_0-\omega_{\varphi}^2}\right).
    \label{hsq1}
\end{equation}
If we introduce the dimensionless parameters
\begin{equation}\label{dim}
\tilde \omega_{\varphi}^2=\frac{\omega_{\varphi}^2}{\omega^2_0} \quad\text{and}\quad \tilde B^2=\frac{g^2B^2_0}{\omega^2_0},
\end{equation}
we can rewrite Eq. \eqref{hsq1} for our non-commutative spacetime as
\begin{equation}
h^2(r)=r^2\Big[\frac{\sqrt{\pi } r^2}{8 \sqrt{\theta }-2 \sqrt{\pi } M r+\sqrt{\pi } r^2}-\frac{\omega^2_{\text{p}}(r)}{\omega^2_0}\Big(1+\frac{ \tilde B^2 }{1-\tilde \omega_{\varphi}^2}\Big)\Big]. \label{eq:hrnew}
\end{equation}
where we have set $g^{tt}$ and $g^{\phi \phi}$ from Eq. \eqref{metric}.
The photon sphere of radius $r_{\text{ph}}$, can be determined by solving the following equation ~\cite{Perlick2015Nov}
\begin{equation}
\frac{d(h^2(r))}{dr}\bigg|_{r=r_{\text{ph}}}=0. \label{eq:con}    
\end{equation}

Now using Eq.~(\ref{eq:hrnew}) in (\ref{eq:con}) we can obtain a non-linear first order differential equation for the photon sphere radius $r_{\text{ph}}$ for our non-commutative BH spacetime in the presence of axion-plasmon effects as:
\begin{equation}
	\begin{aligned}
&\frac{2 \left(\tilde{B}^2-\tilde{\omega }_{\phi }^2+1\right) \omega _p\left(r_{\text{ph}}\right) \left(r_{\text{ph}} \omega _p'\left(r_{\text{ph}}\right)+\omega _p\left(r_{\text{ph}}\right)\right)}{\omega _0^2 \left(\tilde{\omega }_{\phi }^2-1\right)} \\& \hspace{2cm}+\frac{-6 \pi  M r_{\text{ph}}^3+32 \sqrt{\pi } \sqrt{\theta } r_{\text{ph}}^2+2 \pi  r_{\text{ph}}^4}{\left(\sqrt{\pi } r_{\text{ph}} \left(2 M-r_{\text{ph}}\right)-8 \sqrt{\theta }\right){}^2}=0
\end{aligned}
\label{phot_rad}
\end{equation}
The roots of Eq. (\ref{phot_rad}) cannot be obtained analytically because of its non-linearity and presence of mixed coupled terms as well as higher order terms in $r_{ph}$. Nevertheless, we can study it numerically by choosing different forms of $\omega_{\text {p}}(r)$.

Let us now calculate the angular shadow radius which can be defined with the help of the photon sphere radius $r_{ph}$. The angular shadow radius $\alpha_{sh}$ is defined as:
\begin{equation}
\sin^2 \alpha_{sh} = \frac{h^2 (r_{ph})}{h^2 (r_{o})}
\label{ang_shad}
\end{equation}
Using Eq. \eqref{eq:hrnew}, we can express Eq. \eqref{ang_shad} as
\begin{equation}
\sin \alpha_{sh} = \left[ \frac{r_{\text{ph}}^2 \left(\frac{\sqrt{\pi } r_{\text{ph}}^2}{8 \sqrt{\theta }-2 \sqrt{\pi } M r_{\text{ph}}+\sqrt{\pi } r_{\text{ph}}^2}-\frac{\omega _p^2 \left(\frac{\tilde{B}^2}{1-\tilde{\omega }_{\phi }^2}+1\right)}{\omega _0^2}\right)}{r_o^2 \left(\frac{\sqrt{\pi } r_o^2}{8 \sqrt{\theta }-2 \sqrt{\pi } M r_o+\sqrt{\pi } r_o^2}-\frac{\omega _p^2 \left(\frac{\tilde{B}^2}{1-\tilde{\omega }_{\phi }^2}+1\right)}{\omega _0^2}\right)} \right]^{1/2}    
\label{ang_shad1}
\end{equation}
Here, $r_{ph}$ and $r_o$ denotes the photon sphere radius and observer's location respectively. If one assumes that the observer's location lies very far away from the BH, then the approximate BH shadow radius can be found by using Eq. \eqref{ang_shad1} as follows: \cite{Perlick2022Feb}
\begin{equation}
r_{sh} \approx r_o \sin \alpha_{sh},
\label{shad_rad}
\end{equation}
which gives 
\begin{equation}
r_{sh} = r_{ph} \left[ \frac{\sqrt{\pi } r_{\text{ph}}^2}{8 \sqrt{\theta }-2 \sqrt{\pi } M r_{\text{ph}}+\sqrt{\pi } r_{\text{ph}}^2}-\frac{\omega _p^2 \left(\frac{\tilde{B}^2}{1-\tilde{\omega }_{\phi }^2}+1\right)}{\omega _0^2} \right]^{1/2}
\label{shad_rad_gen}
\end{equation}
Note that the shadow radius no longer depends on the observer's location.
We can see that if one reconsiders the limiting case where there is no plasma medium, $\omega_p \to 0$, and where there is no contribution from the axionic and magnetic fields ($\tilde{\omega}_\phi \to 0$ and $\tilde{B}\to 0$ respectively), the trivial case of a Schwarzschild shadow radius ($r_{sh}^{\text{Sch}} = 3\sqrt{3} M$) is obtained at $r_{ph} = 3M$ (photon orbit radius for a Schwarzschild BH).

\subsection{Existence Condition for Circular Orbits: Analytical Estimate}
To ensure the physical existence of shadow-forming photon orbits in the non-commutative BH spacetime with magnetized axion-plasma, we require the plasma frequency squared $\omega_p^2(r)$ to remain positive along photon trajectories. Starting from the analytical trajectory equation \eqref{eq:hrnew}, let
\begin{equation}
\begin{aligned}
    h^2(r) &= r^2 \left[ 
        \frac{\sqrt{\pi} r^2}{8\sqrt{\theta} - 2\sqrt{\pi} M r + \sqrt{\pi} r^2}
         \right. \\& \hspace{2cm} \left. - \frac{\omega_p^2(r)}{\omega_0^2} \left(1 + \frac{\tilde{B}^2}{1-\tilde{\omega}_\varphi^2}\right)
    \right] = C,
    \end{aligned}
\end{equation}
where $C$ is a constant. This is required from the condition \eqref{eq:con}. 
Solving for $\omega_p^2(r)$, we get
\begin{equation}
    \omega_p^2(r) = \omega_0^2
    \left[
        \frac{\sqrt{\pi} r^2}{8\sqrt{\theta} - 2\sqrt{\pi} M r + \sqrt{\pi} r^2}
        - \frac{C}{r^2}
    \right]
    \left(1 + \frac{\tilde{B}^2}{1 - \tilde{\omega}_\varphi^2}\right)^{-1}.
\end{equation}
Requiring $\omega_p^2(r) > 0$ leads to the following necessary condition:
\begin{equation}
    \frac{\sqrt{\pi} r^2}{8\sqrt{\theta} - 2\sqrt{\pi} M r + \sqrt{\pi} r^2} - \frac{C}{r^2} > 0,
\end{equation}
which can be rearranged as the explicit analytical inequality:
\begin{equation}
        \frac{\sqrt{\pi}\, r^4}{8\sqrt{\theta} - 2\sqrt{\pi} M r + \sqrt{\pi} r^2} > C.
    \label{eq:NC_physical_inequality}
\end{equation}
Moreover, one should simultaneously have (for positivity of $\omega_p^2$)
\begin{equation}
1 + \frac{\tilde{B}^2}{1 - \tilde{\omega}_\varphi^2} > 0.
\label{eq:NC_physical_inequality2}
\end{equation}
To satisfy this condition, \(\tilde{\omega}_\varphi\) must lie within the range
\begin{equation}
\tilde{\omega}_\varphi < -\sqrt{1 + \tilde{B}^2} \,\, \text{or} \,\, -1 < \tilde{\omega}_\varphi < 1 \,\, \text{or} \,\, \tilde{\omega}_\varphi > \sqrt{1 + \tilde{B}^2}
\end{equation}
irrespective of the sign of \(\tilde{B}\).

To sum up, the constraint \eqref{eq:NC_physical_inequality} determines for which parameter values of non-commutativity, shadow-forming orbits are permitted. If this inequality is violated for all $r$, then no shadow can form. In the standard Schwarzschild limit (absence of non-commutative contribution $\theta \to 0$), Eq.~\eqref{eq:NC_physical_inequality} reduces to the standard $r^3/(r-2M) > C$ condition obtained by Perlick et al. (Eq. (36) of \cite{Perlick2015Nov}). This analytical inequality provides a clear criterion which links the spacetime geometry and plasma properties, thereby offering a direct way to estimate how varying plasma density or non-commutative effects constrain or even suppress the existence of shadow-forming photon orbits in these BH spacetimes. 

Also, the constraint \eqref{eq:NC_physical_inequality2} shows that the axion/magnetic parameter \(\tilde{\omega}_\varphi\) is only allowed within the open interval \((-1, 1)\) or outside the broad forbidden region \([-\sqrt{1+\tilde{B}^2}, \sqrt{1+\tilde{B}^2}]\). For values of \(\tilde{\omega}_\varphi\) close to \(\pm1\) or for sufficiently large couplings where \(|\tilde{\omega}_\varphi|\) exceeds \(\sqrt{1+\tilde{B}^2}\), the effective plasma-magnetic term becomes negative or divergent. This suppresses shadow formation and leads to an unphysical situation.

For all our choices of $\tilde{B}^2 = 0.1, 0.3, 0.5$ in our subsequent analysis, the positivity condition for the effective plasma-axion-magnetic term is satisfied for all $\tilde{\omega}_\varphi$ in the range $|\tilde{\omega}_\varphi| < 1$, which covers the physically relevant scenarios considered in this work. The constraint only excludes a very narrow interval near $|\tilde{\omega}_\varphi| = 1$ (specifically, $1 < |\tilde{\omega}_\varphi| < \sqrt{1+\tilde{B}^2}$ for each $\tilde{B}^2$), which we have not considered in our analysis.

It is worthy of mention that there exist other equivalent techniques to analytically examine the existence of BH shadows, one of such techniques was studied by Vertogradov and \"Ovg\"un \cite{Vertogradov2024Jul} (and references therein), based on geometrical deformation using extended gravitational decoupling technique. Some of the other methods relating BH shadows with various other properties like plasma effects, rotation and charge are discussed widely in the works of \cite{Perlick2022Feb,Ovgun2018Oct,Tsukamoto2018Mar}. Another method was discussed by Tsupko using deformation of shadow to analytically estimate the BH spin \cite{Tsupko2017May}. A related work concerning spin constraints and other deformation parameters using BH shadow was also carried out by Tsukamoto et al \cite{Tsukamoto2014Jun}.

\section{Photon Sphere Radius}
\label{sec3}
In this section, we shall discuss how in the presence of a plasma medium, the effective motion of photons is affected due to the dispersive properties of the plasma. In this context, we shall adopt two specific cases. First, we assume a homogeneous plasma, which is characterized by a constant electron number density and plasma frequency, and hence a uniform refractive index. Secondly, we shall consider an inhomogeneous plasma, characterized by a power-law varying plasma frequency and hence a spatially varying refractive index. To be specific, we want to see how the photon radius depends on the particular nature of surrounding plasma medium.

\subsection{Homogeneous Plasma} 
\label{sec3a}
\begin{figure*}[htb]
\centerline{\includegraphics[scale=0.45]{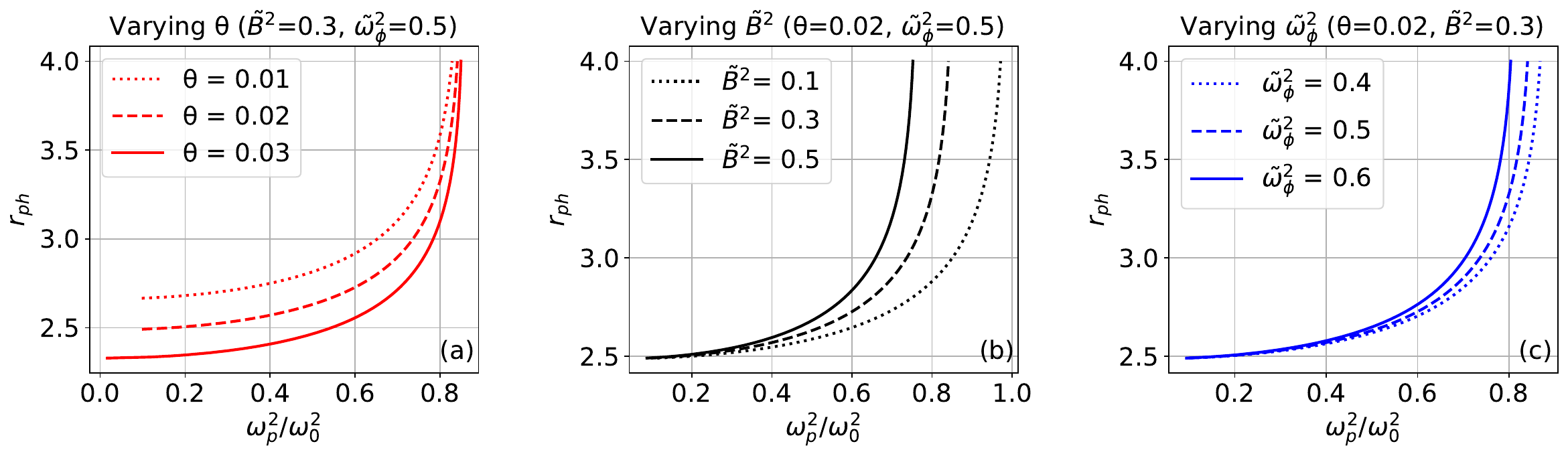}}
\caption{Variation of photon sphere radius with the plasma frequency for various combinations of the parameters for a homogeneous plasma}
\label{homo1}
\end{figure*}

In case the non-commutative BH is surrounded by a homogeneous plasma characterized by a fixed plasma frequency i.e. $\omega_p^2= \text{const.}$, the exact analytical expression for the radius of photon sphere from Eq. \eqref{phot_rad} can be hard to obtain. This is because, the surrounding plasma also constitutes the axionic field frequency $\tilde{\omega}_\phi$ as well as a magnetic field $\tilde{B}$, coupled with the plasma frequency $\omega_{p}$, making them interdependent. Therefore, the photon sphere radius should be calculated numerically. In fact to be more specific, we shall plot the same with respect to the plasma frequency by setting different values of our three parameters of dependency, i.e., the non-commutative parameter $\theta$, axion frequency $\tilde{\omega}_{\phi}^2$,and the magnetic field $\tilde{B}^2$. For simplicity, we shall work with units where $M=1$. The numerical results are plotted in Fig.~\ref{homo1}. In the figure, we can see how the photon sphere radius varies with the plasma frequency, for any three values of the aforementioned model parameters. In Fig.~\ref{homo1} (a), for fixed values of $\tilde{\omega}_\phi^2$ and magnetic field $\tilde{B}^2$, the photon sphere radius rises exponentially with the increase in plasma frequency $\omega_{p}^2/\omega_0^2$. One can see, the photon sphere radius starts with smaller values for larger values of $\theta$. This implies that as $\theta$ increases, the shadow radius decreases. In Fig. \ref{homo1} (b) and (c), however we see a contrasting scenario with (a). In the former, keeping $\theta = 0.02$ and $\tilde{\omega}_{\phi}^2 = 0.5$, we can see an exponential increase in photon sphere radius with increasing values of $\omega_{ph}^2/\omega_0^2$ for different values of $\tilde{B}^2$ with $\theta$ and $\tilde{\omega}_\phi^2$ fixed, but for all the cases, the radius starts with the same value. This implies that, in the vacuum case $\omega_{p} \to 0$, there is no dependence of photon sphere radius with varying $\tilde{B}^2$. The slope of the curve however increases with increasing values of $\tilde{B}^2$. In the latter case, the same behaviour is apparent with the variation with the axion frequency $\omega_\phi^2$, while keeping $\theta$ and $\tilde{B}^2$ fixed. But, the slope of the curves vary moderately with changing values of axion frequency.

\begin{figure*}[htb]
\centerline{\includegraphics[scale=0.45]{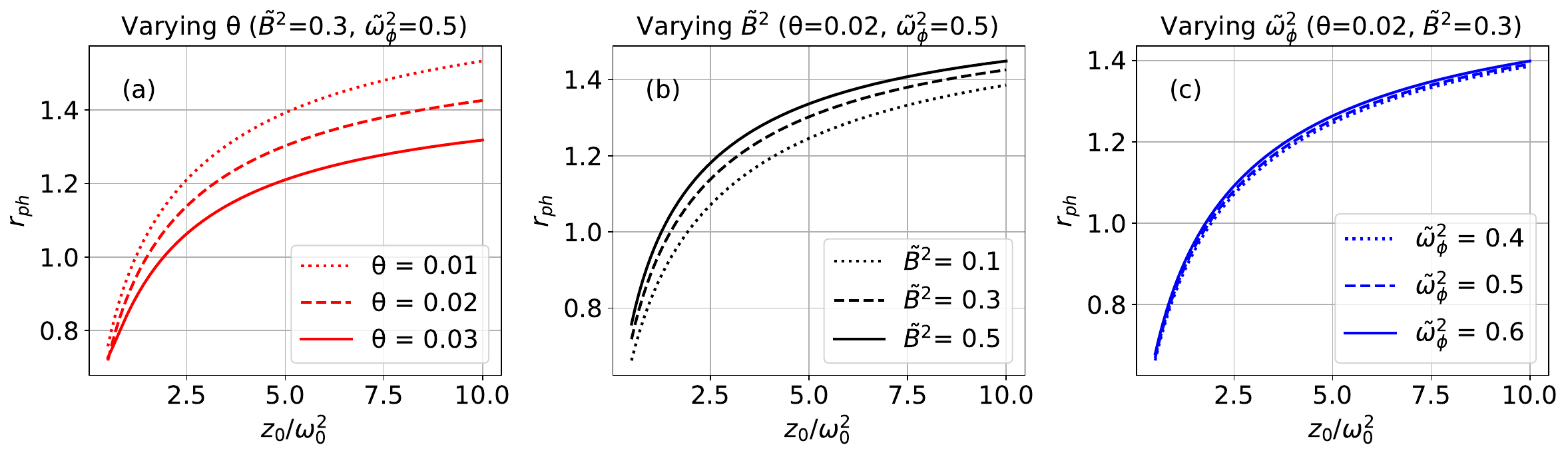}}
\caption{Variation of photon sphere radius with the plasma frequency for various combinations of the parameters for an inhomogeneous plasma}
\label{inhomo1}
\end{figure*}

\subsection{Inhomogeneous plasma with $\omega^2_{p}(r)=\kappa/r^q$}
Next, let us investigate the dependence of an inhomogeneous plasma on the radii of photon spheres while including the axionic contribution. In this case, we may consider that plasma frequency as a radially dependent function following a power-law form as: \cite{Rog:2015a,Er2017aa}
\begin{equation}\label{inhomo_plasma_freq}
\omega^2_{p}(r)=\frac{\kappa}{r^q}.
\end{equation}
In this context, $\kappa$ and $q$ are independent free parameters. The values of these parameters in the present work shall be fixed to $q=1$ and $\kappa$ as a coupling constant \cite{Rog:2015a}. The parameter $q$ in fact quantifies how the intensity of plasma effect falls over radial distances. For instance, if one sets $q = 2$, the plasma frequency falls off like an inverse square law and would signify a faster dilution of plasma effects as compared to $q = 1$. Therefore, $q = 1$ case would provide a clearer impact of the presence of the plasma medium on the BH optical properties. Once again, we seek numerical solutions of the photon sphere radii by solving Eq. \eqref{phot_rad} for the case of inhomogeneous plasma. After solving Eq. \eqref{phot_rad}, the photon sphere radii are plotted against the coupling constant of inhomogeneous plasma $\kappa / \omega_0^2$. From Fig. \ref{inhomo1} (a), for a fixed value of $\tilde{B}^2 = 0.3$ and $\tilde{\omega}_\phi^2 = 0.5$ we can see that the variation of the photon sphere radius increases initially, then tends to flatten out as the coupling constant $\kappa/ \omega_0^2$ increases. Also, as the non-commutative parameter decreases, the slope of the curve increases noticeably, signifying increase in value of $r_{ph}$ as $\theta$ decreases. Note that in the vacuum limit, $\kappa \to 0$, the photon sphere radius reduces to the same initial value, irrespective of the value of $\theta$. Next, in Figs. \ref{inhomo1} (b) and (c), we see the same behaviour, except in a way how the photon sphere radius vary with respect to changing $\tilde{B}^2$ (see Fig. \ref{inhomo1}(b)). In other words, towards the vacuum limit, the photon sphere radius tends to have different initial values and these values increases with increasing $\tilde{B}^2$. The overall slope of the curves has almost the same magnitude but with smaller deviations. In Fig. \ref{inhomo1} (c), with respect to varying axion frequency $\tilde{\omega}_{\phi}^2$, the trend of variation of photon sphere radius is almost similar to the previous cases, but the effect of the axion field is minimal and no significant dependence on the slope of the curve is prominent here.

\section{Shadows}
\label{sec4}
\subsection{Homogeneous Plasma}
\label{subsecshad1}
\begin{figure*}[htb]	\centerline{\includegraphics[scale=0.45]{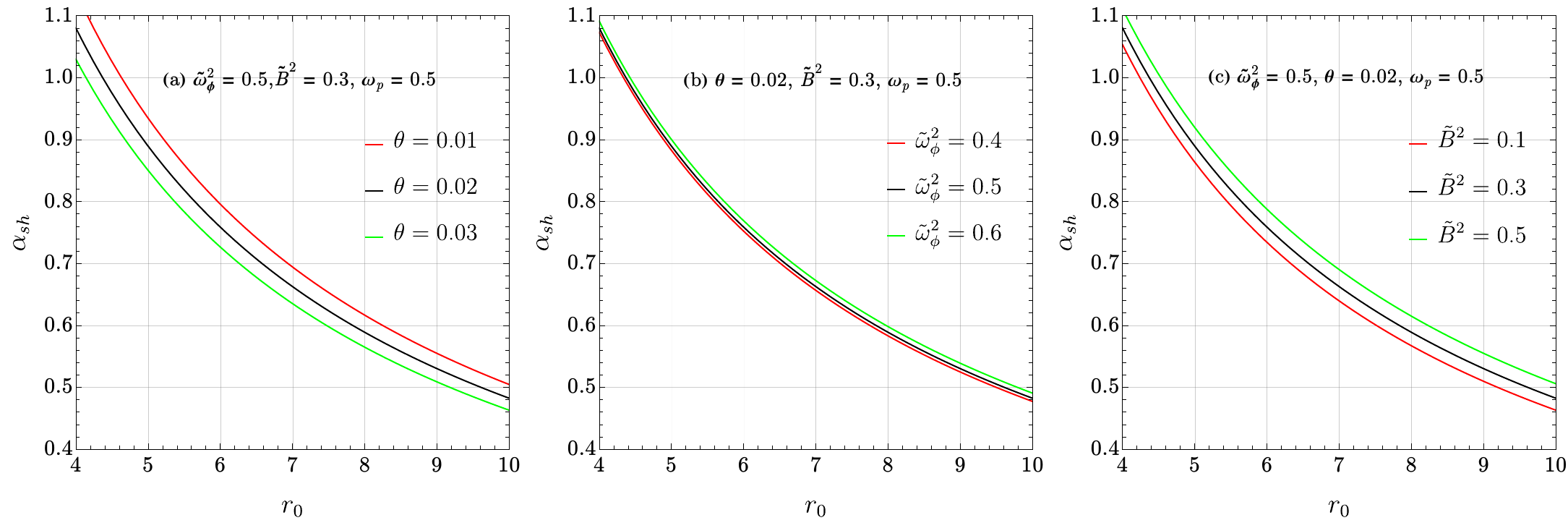}}
	\caption{Variation of angular shadow with the observer's position for various combinations of the parameters for a homogeneous plasma}
	\label{ang_shad_homo1}
\end{figure*}

Now let us discuss the behaviour of shadow cast by the non-commutative BH in the presence of plasma. To start with, we calculate the angular shadow through Eq. \eqref{ang_shad1}. In homogeneous plasma case, we can have:
\begin{equation}
	\sin \alpha_{sh} = \left[ \frac{r_{\text{ph}}^2 \left(\frac{\sqrt{\pi } r_{\text{ph}}^2}{8 \sqrt{\theta }-2 \sqrt{\pi } M r_{\text{ph}}+\sqrt{\pi } r_{\text{ph}}^2}-\frac{\omega _p^2 \left(\frac{\tilde{B}^2}{1-\tilde{\omega }_{\phi }^2}+1\right)}{\omega _0^2}\right)}{r_o^2 \left(\frac{\sqrt{\pi } r_o^2}{8 \sqrt{\theta }-2 \sqrt{\pi } M r_o+\sqrt{\pi } r_o^2}-\frac{\omega _p^2 \left(\frac{\tilde{B}^2}{1-\tilde{\omega }_{\phi }^2}+1\right)}{\omega _0^2}\right)} \right]^{1/2}    
	\label{ang_shad_homo}
\end{equation}
where $\omega_p$ is treated as a constant. Fig. \ref{ang_shad_homo1} shows the numerical plot of the angular shadow with respect to observer's position. In Fig. \ref{ang_shad_homo1}(a), the axion frequency \(\tilde{\omega}_{\phi}\) and \(\tilde{B}^2\) are kept fixed at 0.5 and 0.3, respectively, while the constant homogeneous plasma frequency is set to \(\omega_p = 0.5\). We have solved for the photon sphere radius $r_{ph}$ from the condition \eqref{phot_rad}, for each case and used those in Eq. \eqref{ang_shad_homo}. It is seen that the angular shadow \(\alpha_{sh}\) decreases asymptotically with increasing observer distance, with the variation noticeably influenced by the non-commutative parameter \(\theta\). As \(\theta\) increases, the angular shadow starts from smaller values, while the slope of the curves remains unchanged. Figs. \ref{ang_shad_homo1}(b) and (c) also exhibits analogous behavior. In Fig. \ref{ang_shad_homo1}(b), the dependence of the angular shadow on the axion frequency is relatively weak provided \(\theta = 0.02\), \(\tilde{B}^2 = 0.3\), and \(\omega_p = 0.5\) are held constant. In Fig. \ref{ang_shad_homo1}(c), under the conditions \(\theta = 0.02\), \(\tilde{\omega}_\phi^2 = 0.3\), and \(\omega_p = 0.5\), the angular shadow continues to decrease with observer distance, yet it tends to converge to similar values for observers positioned at nearer regions of the BH. 

\begin{figure*}[htb]
	\centerline{\includegraphics[scale=0.42]{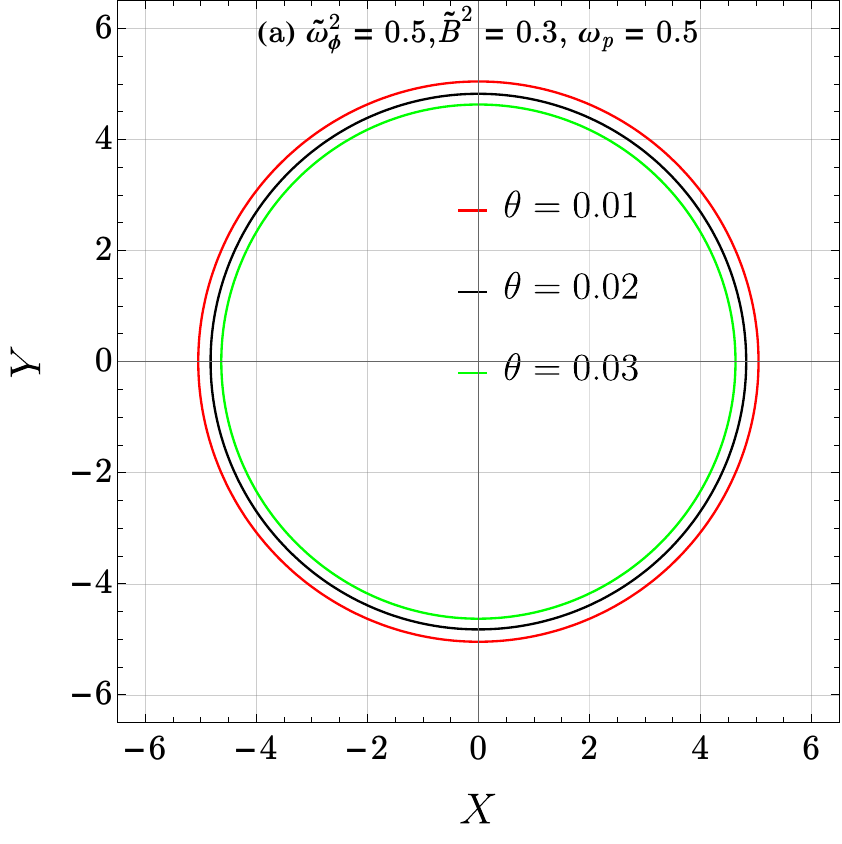} \hspace{0.1cm} \includegraphics[scale=0.42]{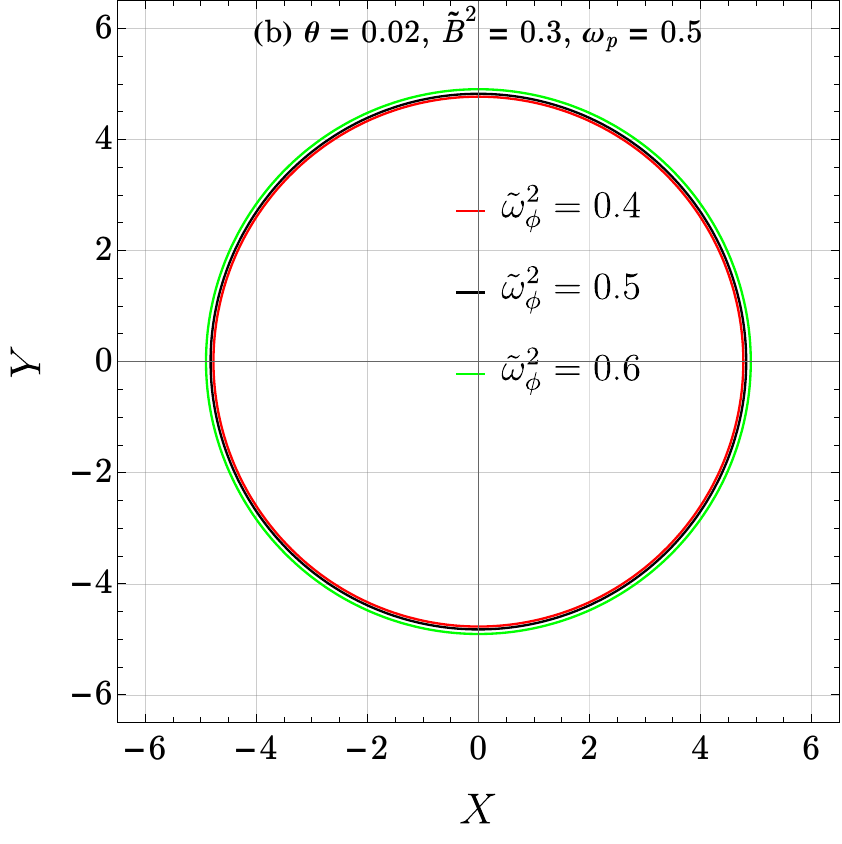} \hspace{0.1cm} \includegraphics[scale=0.42]{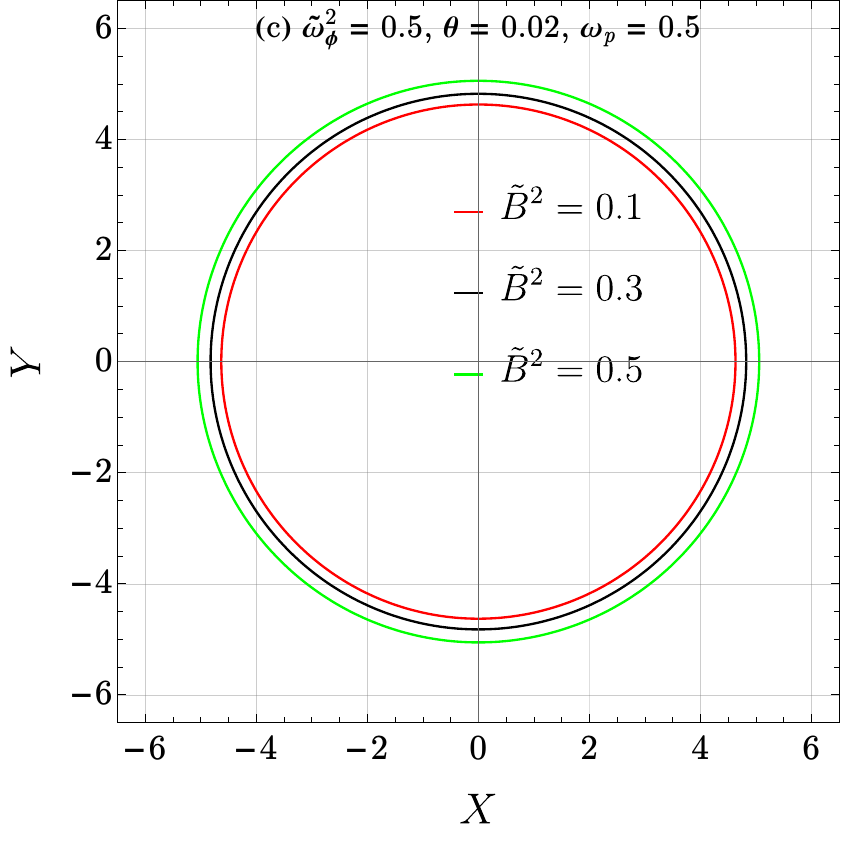}}
	\caption{Shadow images for various combinations of the parameters for a homogeneous plasma}
	\label{shad_homo1}
\end{figure*}

The shadow radius can be calculated from Eq. \eqref{shad_rad_gen} as 
\begin{equation}
r_{sh} = r_{ph} \left[ \frac{\sqrt{\pi } r_{\text{ph}}^2}{8 \sqrt{\theta }-2 \sqrt{\pi } M r_{\text{ph}}+\sqrt{\pi } r_{\text{ph}}^2}-\frac{\omega _p^2 \left(\frac{\tilde{B}^2}{1-\tilde{\omega }_{\phi }^2}+1\right)}{\omega _0^2} \right]^{1/2}
\label{shad_rad_exp_homo}
\end{equation}
The expression remains same as that of Eq. \eqref{shad_rad_gen} except here $\omega_p^2$ is a constant.
Using Eq. \eqref{shad_rad_exp_homo}, we plotted of the shadows of the BH in the Cartesian plane, as shown in Fig. \ref{shad_homo1}. In Fig. \ref{shad_homo1}(a), it is clear that the non-commutative parameter $\theta$, while having fixed values for $\tilde{\omega}_{\phi}^2$ and $\tilde{B^2}$ and setting $\omega_p$ to 0.5, clearly influences the shadow size. In particular, an increase in $\theta$ results in a smaller shadow. A similar effect is observed while varying $\tilde{B}^2$, where an increase in the magnetic field causes an expansion in the shadow radius, as can be seen from Fig. \ref{shad_homo1}(c). In contrast, changes in $\tilde{\omega}_{\phi}^2$ have a very minimal effect on the size of the shadows, as is observed from Fig. \ref{shad_homo1} (b).
\subsection{Inhomogeneous Plasma}
\label{subshassec2}
Next, in the inhomogeneous plasma case setting $\omega_{p}^2 = \kappa/r^q$, where $q = 1$ in Eq. \eqref{ang_shad1} we can obtain the angular shadow as:
\begin{equation}
\sin \alpha_{sh} = \left[\frac{r_{\text{ph}}^2 \left(\frac{\sqrt{\pi } r_{\text{ph}}^2}{8 \sqrt{\theta }-2 \sqrt{\pi } M r_{\text{ph}}+\sqrt{\pi } r_{\text{ph}}^2}-\frac{\kappa ^2 \left(\frac{\tilde{B}^2}{1-\tilde{\omega }_{\phi }^2}+1\right)}{\omega _0^2 r_{\text{ph}}^2}\right)}{r_o^2 \left(\frac{\sqrt{\pi } r_o^2}{8 \sqrt{\theta }-2 \sqrt{\pi } M r_o+\sqrt{\pi } r_o^2}-\frac{\kappa ^2 \left(\frac{\tilde{B}^2}{1-\tilde{\omega }_{\phi }^2}+1\right)}{\omega _0^2 r_o^2}\right)}\right]^{1/2}
\label{ang_shad_inhomo}
\end{equation}

\begin{figure*}[htb]
	\centerline{\includegraphics[scale=0.45]{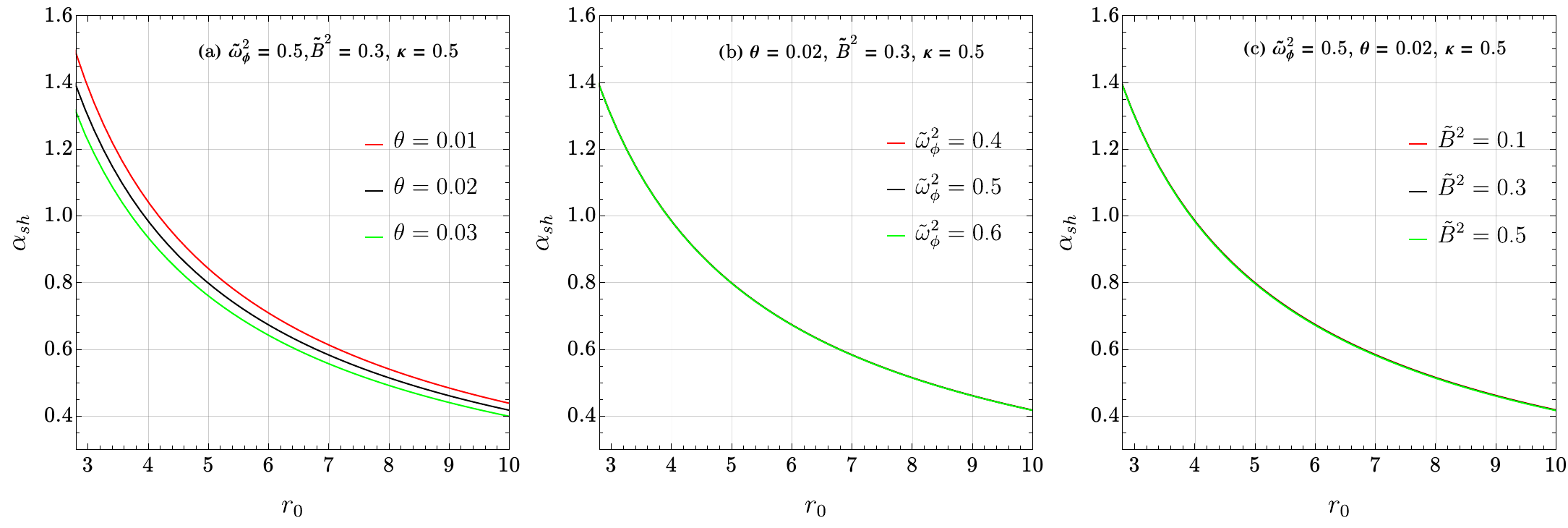}}
	\caption{Variation of angular shadow with the observer's position for various combinations of the parameters for an inhomogeneous plasma}
	\label{ang_shad_inhomo1}
\end{figure*}

Eq. \eqref{ang_shad_inhomo} is analyzed by plotting for various combinations of the parameters, like we did in the previous subsection for the homogeneous plasma case. The key distinction in this analysis lies in the inclusion of the inhomogeneous plasma coupling parameter, \(\kappa\). For the purpose of this work, we have adopted a constant value of \(\kappa = 0.5\), though it is important to note that \(\kappa\) may assume any arbitrary positive values in general. 
Fig. \ref{ang_shad_inhomo1} (a) shows the variation of the angular shadow as a function of the observer's position, for several different values of the non-commutative parameter \(\theta\). In this case, we have set \(\tilde{\omega}_\phi^2 = 0.5\) and \(\tilde{B}^2 = 0.3\). It is observed that the angular shadow decreases as the observer moves farther from the BH. Notably, the dependence on \(\theta\) becomes evident: as the observer approaches the BH, the angular shadow tends to increase for smaller values of the non-commutative parameter \(\theta\). 
In Fig. \ref{ang_shad_inhomo1} (b), the variation of the angular shadow is shown for different values of the axion frequency \(\tilde{\omega}_\phi^2\), while maintaining fixed values for \(\theta\) and \(\tilde{B}^2\). The overall trend is qualitatively similar to the previous cases, with a very feeble dependence on the axion frequency. A similar observation can be made when considering the dependence on varying \(\tilde{B}^2\) values, as displayed in Fig. \ref{ang_shad_inhomo1} (c). 
From these analyses, it is clear that neither the axion frequency \(\tilde{\omega}_\phi^2\) nor the magnetic field \(\tilde{B}^2\) exerts a significant influence on the angular shadow within the context of an inhomogeneous plasma. These results points towards the conclusion that the primary factor influencing the angular shadow in this scenario is the observer's position, with only noticeable effects arising from variations in the non-commutative parameter \(\theta\). The dependence on the axion frequency and magnetic field however is quite subtle in the case of inhomogeneous plasma. This is one of the important result of this analysis.

The shadow radius in the case of inhomogeneous plasma can be calculated from Eq. \eqref{shad_rad_gen} as 
\begin{equation}
r_{sh} = r_{ph} \left[\frac{\sqrt{\pi } r_{\text{ph}}^2}{8 \sqrt{\theta }-2 \sqrt{\pi } M r_{\text{ph}}+\sqrt{\pi } r_{\text{ph}}^2}-\frac{\kappa ^2 \left(\frac{\tilde{B}^2}{1-\tilde{\omega }_{\phi }^2}+1\right)}{\omega _0^2 r_{\text{ph}}^2} \right]^{1/2}
\label{shad_rad_exp_inhomo}
\end{equation}

\begin{figure*}[htb]
	\centerline{\includegraphics[scale=0.45]{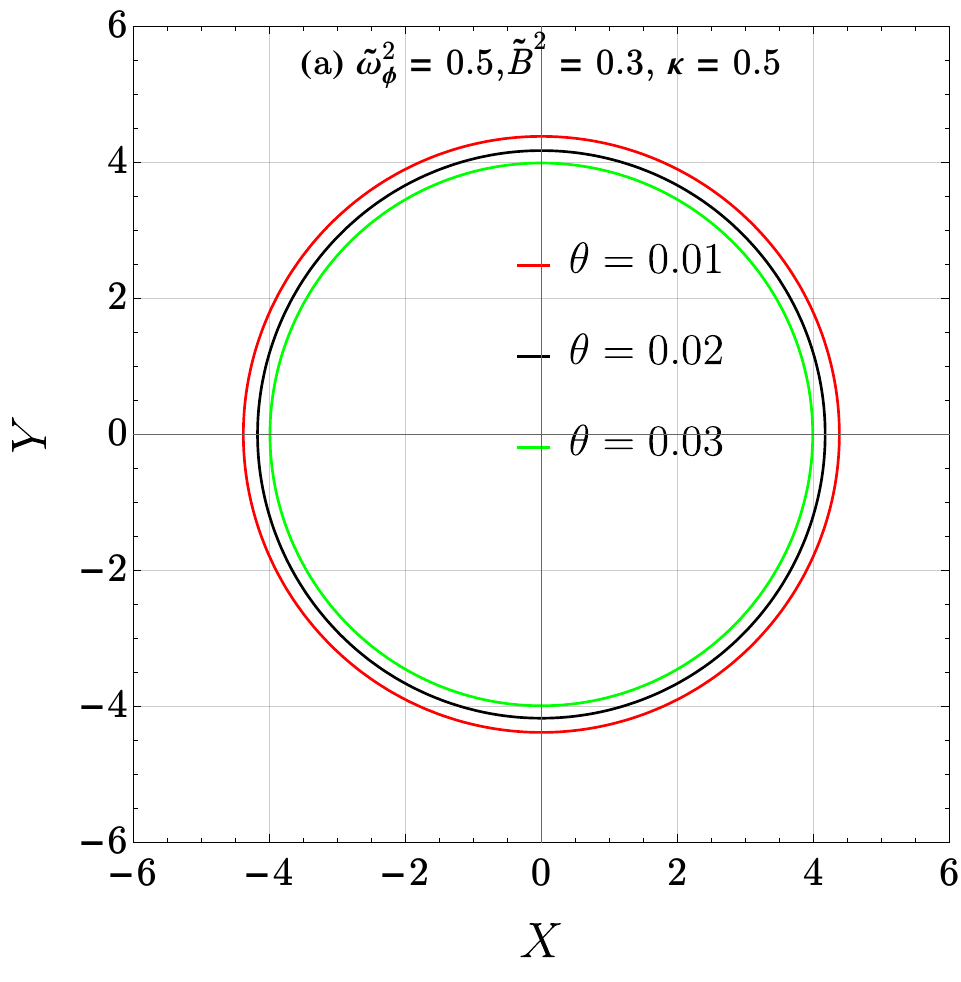}}	\centerline{\includegraphics[scale=0.45]{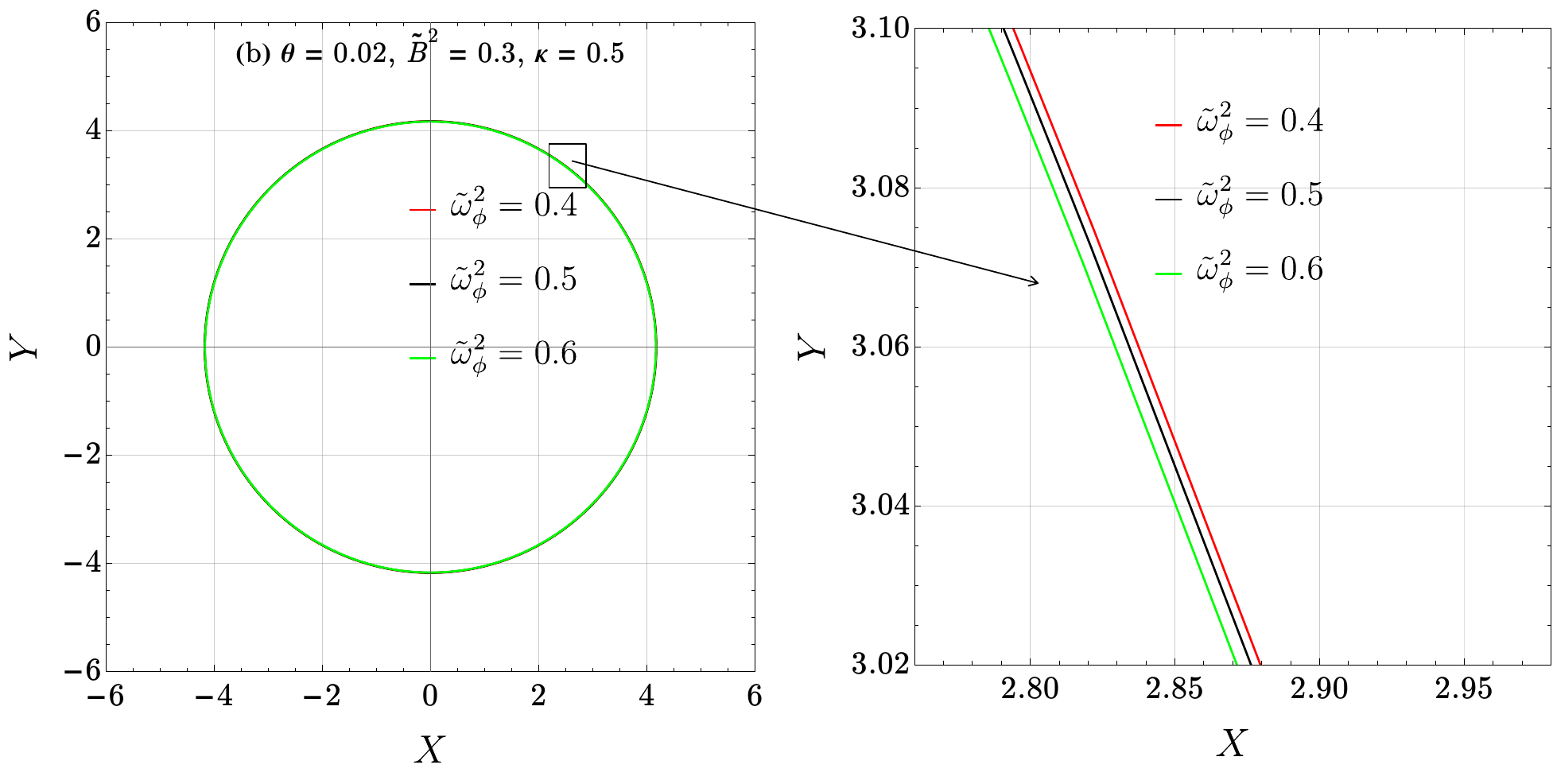}} \centerline{\includegraphics[scale=0.45]{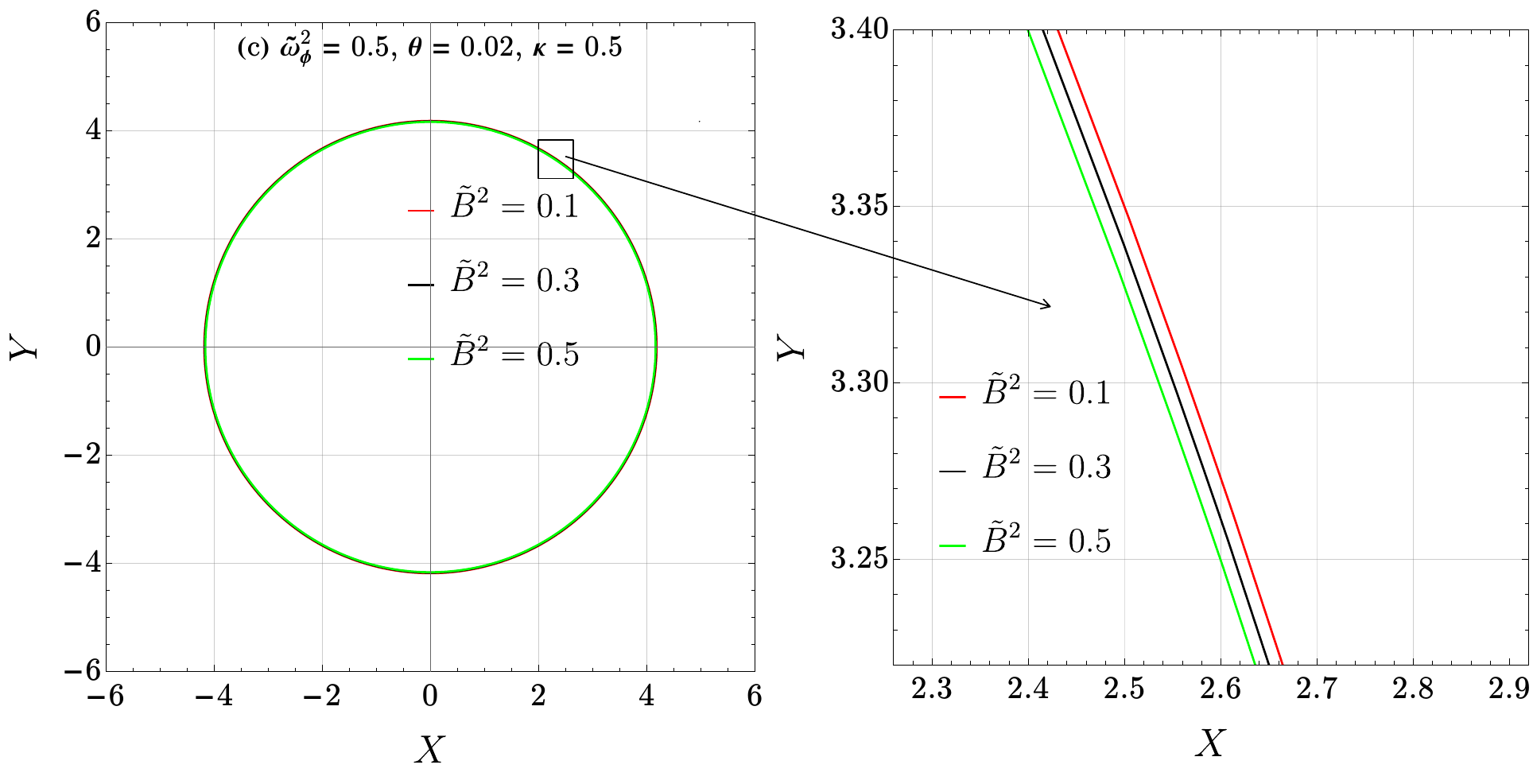}}
	\caption{Shadow images for various combinations of the parameters for an in homogeneous plasma}
	\label{shad_inhomo1}
\end{figure*}

Using Equation \eqref{shad_rad_exp_inhomo}, we have generated a plot of the shadows in the Cartesian plane, as portrayed in Fig. \ref{shad_inhomo1}. In Fig. \ref{shad_inhomo1}(a), it is evident that the non-commutative parameter $\theta$, while maintaining fixed values of $\tilde{\omega}_{\phi}^2$ and $\tilde{B^2}$ and setting $\kappa$ to 0.5, uniquely influences the size of the shadows. Consistent with the results in the homogeneous plasma scenario in subsection \ref{subsecshad1}, an increase in the non-commutative parameter leads to a reduction in shadow size. Furthermore, Figs. \ref{shad_inhomo1} (b) and (c) exhibits a subtle dependence on the variations of $\tilde{\omega}_{\phi}$ and $\tilde{B}^2$, respectively. A closer look at the zoomed in figures of (b) and (c) reveals that an increase in $\tilde{\omega}_{\phi}^2$ leads to a decrease in the shadow radius, and similarly, an increase in $\tilde{B}^2$ also results in a smaller shadow radius.



\section{Observational Constraints}
\label{sec5}
In the strong-gravity regime, the EHT's most recent horizon-scale image of Sgr A$^*$ \cite{EventHorizonTelescopeCollaboration2022May} offers an ideal platform to test theories of gravity and fundamental physics.  Because Sgr A$^*$ is so very near to us, estimating its mass, distance, and, consequently, its mass-to-distance ratio, becomes very much easier.  Compared to M87*, whose mass is not strictly restricted by the stellar dynamics observations, this poses itself as a significant benefit.  The mass of the Sgr A$^*$ ($O(10^6)M_\odot$) is much smaller than that of the M87$^*$ ($O(10^9)M_\odot$), opening up a whole new way of looking at fundamental physics in the strong curvature regime.  The Sgr A's shadow can also offer more rigorous limits on fundamental parameters than the M87's, using the same reasoning \cite{Vagnozzi2023Jul}.

The EHT has determined the fractional deviation between the predicted shadow radius $r_s$ and the shadow radius of a Schwarzschild BH $r_{sch} = 3\sqrt{3} M$, given by:
\begin{equation}
\delta \equiv \frac{r_s}{r_{sch}} - 1 = \frac{r_{s}}{3\sqrt{3} M} - 1.
\label{del_shad}
\end{equation}
The Keck and VLTI set the following constraints on $\delta$ as: \cite{Do2019Jul,Abuter2020Apr,Vagnozzi2023Jul}
\[ \delta = -0.04^{+0.09}_{-0.10} \text{  (Keck) \quad  and  \quad } 
\delta = -0.08^{+0.09}_{-0.09} \text{  (VLTI)} \]
For a convenient analysis, we may calculate the average value of $\delta$ estimated from the Keck and VLTI observations since these values are uncorrelated as they are originated from independent measurements. Thus the average value of $\delta$ is \cite{Do2019Jul,Abuter2020Apr,Vagnozzi2023Jul}
\begin{equation}
\delta \approx -0.060 \pm 0.065.
\label{av_del}
\end{equation}
If ones assumes a Gaussian posterior distribution, the confidence intervals for $\delta$ are
\begin{equation}
 -0.125 \lesssim \delta  \lesssim 0.005 \quad (1 \sigma) \text{  and }  -0.19 \lesssim \delta  \lesssim 0.07 \quad(2 \sigma).
 \label{const_aver}
\end{equation}
Extracting $r_s$ from Eq. \eqref{del_shad}, we get 
\begin{equation}
\frac{r_s}{M} = 3 \sqrt{3} (\delta + 1),
\label{rs_m}
\end{equation}
which when combined with the constraints Eq. \eqref{const_aver}, gives
\begin{equation}
4.55 \lesssim r_s/M \lesssim 5.22, \,(1\sigma) \text{ and } 
4.21 \lesssim r_s/M \lesssim 5.56, \,(2 \sigma).
\label{rs_const}
\end{equation}

\renewcommand{\arraystretch}{1.2}
\begin{table*}[htb]
\begin{tabular}{|c|c|c|}
\hline
\multicolumn{3}{|c|}{Homogeneous Plasma}                                                                                                  \\ \hline
\multicolumn{1}{|c|}{Parameter}                 & \multicolumn{1}{c|}{$1\sigma$}                & $2\sigma$                               \\ \hline
\multicolumn{1}{|c|}{$\theta$}                  & \multicolumn{1}{c|}{$0 < \theta \le 0.01448$} & $0 < \theta \le 0.0322$                 \\ \hline
\multicolumn{1}{|c|}{$\tilde{\omega}_{\phi}^2$} & \multicolumn{1}{c|}{$--$}                     & $0 < \tilde{\omega}_{\phi}^2 \le 0.942$ \\ \hline
\multicolumn{1}{|c|}{$\tilde{B}^2$}             & \multicolumn{1}{c|}{$ -- $}                   & $ 0 < \tilde{B}^2 < 0.4788$             \\ \hline
\end{tabular}
\vspace{0.5cm}
\begin{tabular}{|c|c|c|}
\hline
\multicolumn{3}{|c|}{Inhomogeneous Plasma}                                                                                                       \\ \hline
\multicolumn{1}{|c|}{Parameter}                 & \multicolumn{1}{c|}{$1\sigma$}                      & $2\sigma$                                \\ \hline
\multicolumn{1}{|c|}{$\theta$}                  & \multicolumn{1}{c|}{$0.02750 < \theta \le 0.05881$} & $0 .0122 < \theta \le 0.06$              \\ \hline
\multicolumn{1}{|c|}{$\tilde{\omega}_{\phi}^2$} & \multicolumn{1}{c|}{$--$}                           & $0 < \tilde{\omega}_{\phi}^2 \le 0.6262$ \\ \hline
\multicolumn{1}{|c|}{$\tilde{B}^2$}             & \multicolumn{1}{c|}{$ 0 < \tilde{B}^2 < 0.2402$}    & $ 0 < \tilde{B}^2 < 0.4738$              \\ \hline
\end{tabular}
\caption{The constraints on the model parameters at 1$\sigma$ and 2$\sigma$ confidence intervals with respect to the Sgr* shadow data.}
\label{tab1}
\end{table*}

\begin{figure*}[htb]
	\centerline{\includegraphics[scale=0.3]{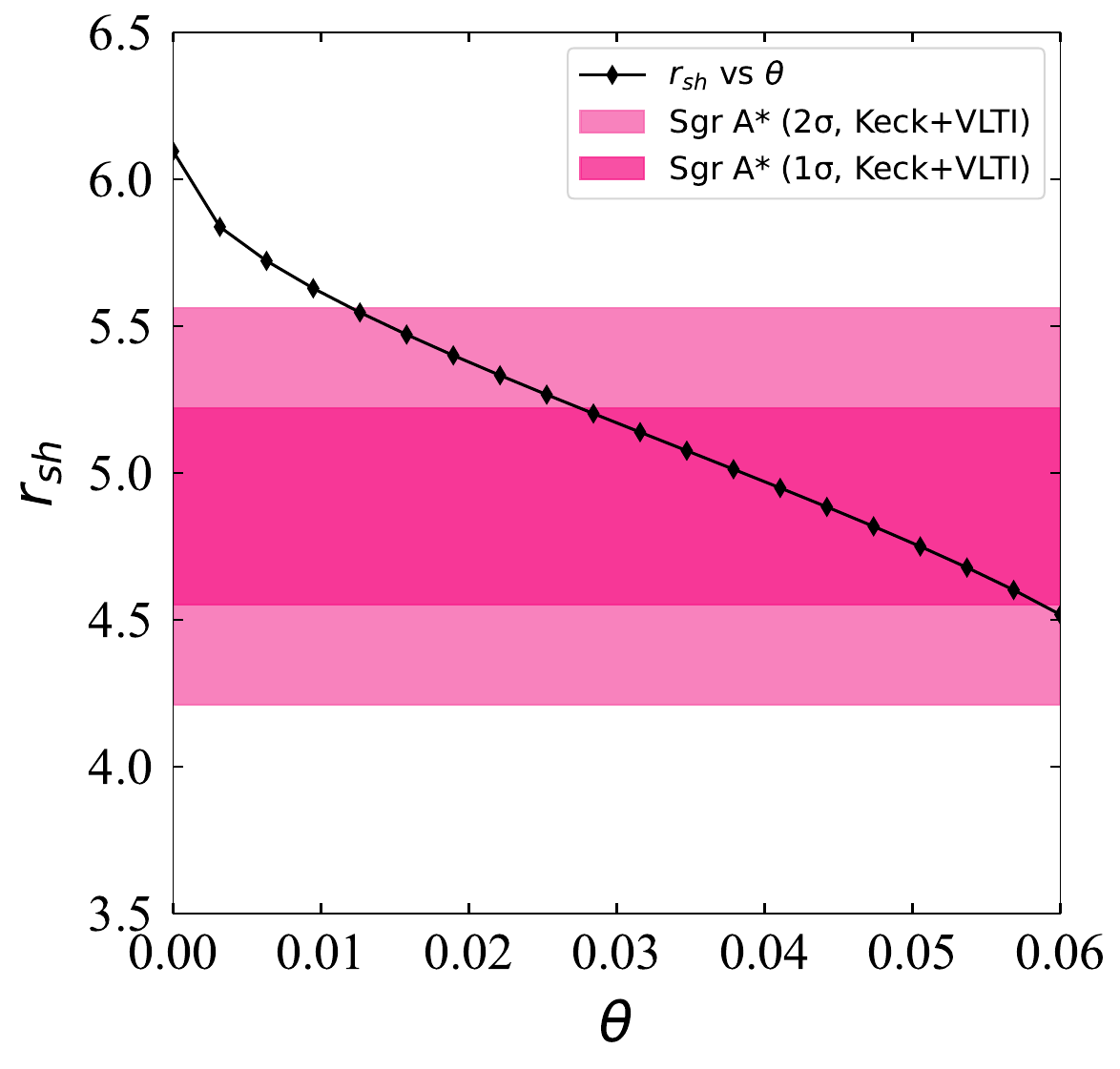}\hspace{0.1cm}\includegraphics[scale=0.3]{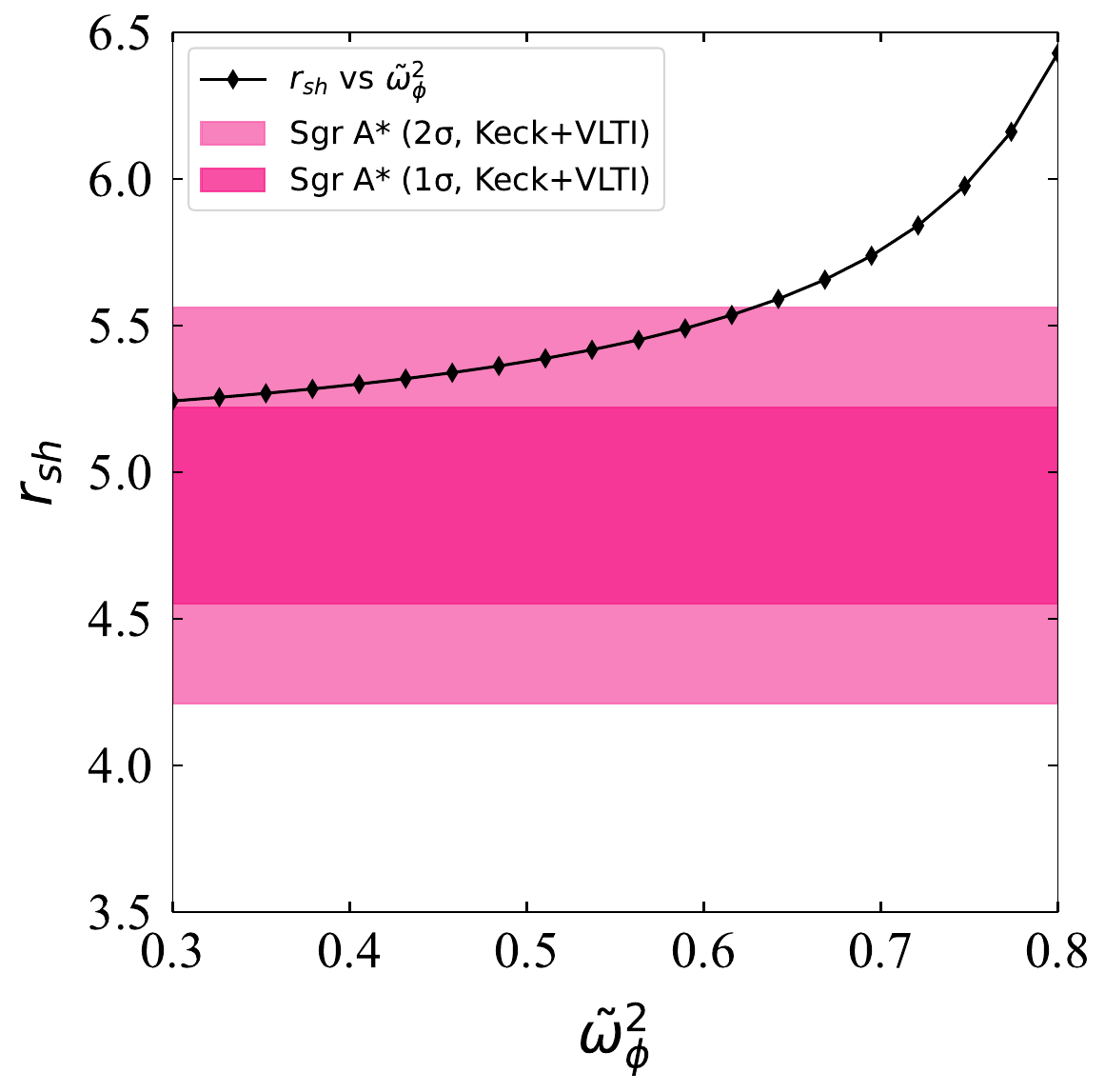} \hspace{0.1cm}\includegraphics[scale=0.3]{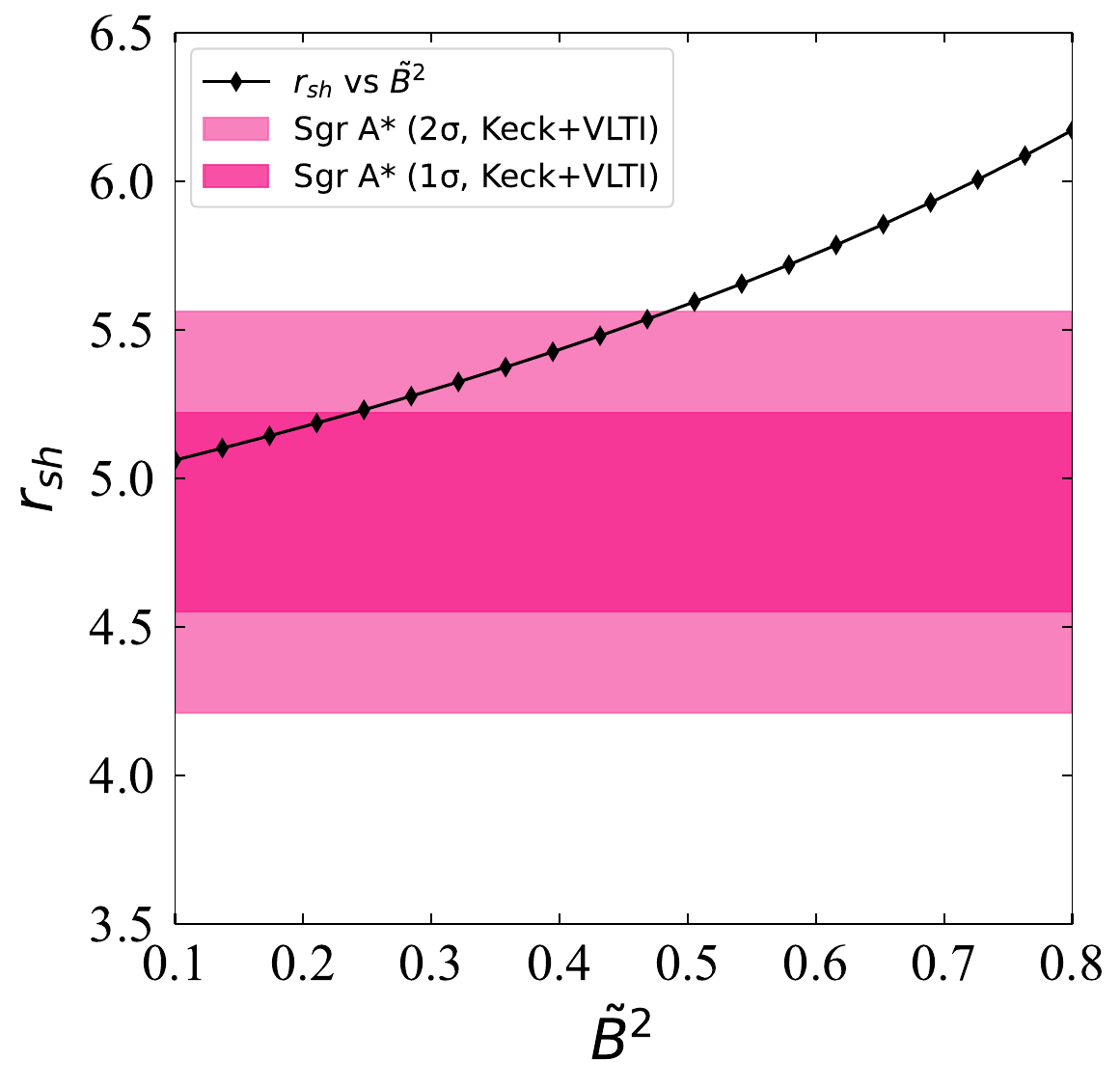}}
	\caption{Variation of shadow radius with the model parameters for homoegeneous plasma case with respect to the observational data. Here we have set $\omega_0^2 =1$, $\tilde{\omega}_{\phi}^2 = 0.5$ and $\tilde{B}^2 = 0.3$. Here we have set the observer at $r_o = 75$.}
\end{figure*}

\begin{figure*}[htb]
	\centerline{\includegraphics[scale=0.3]{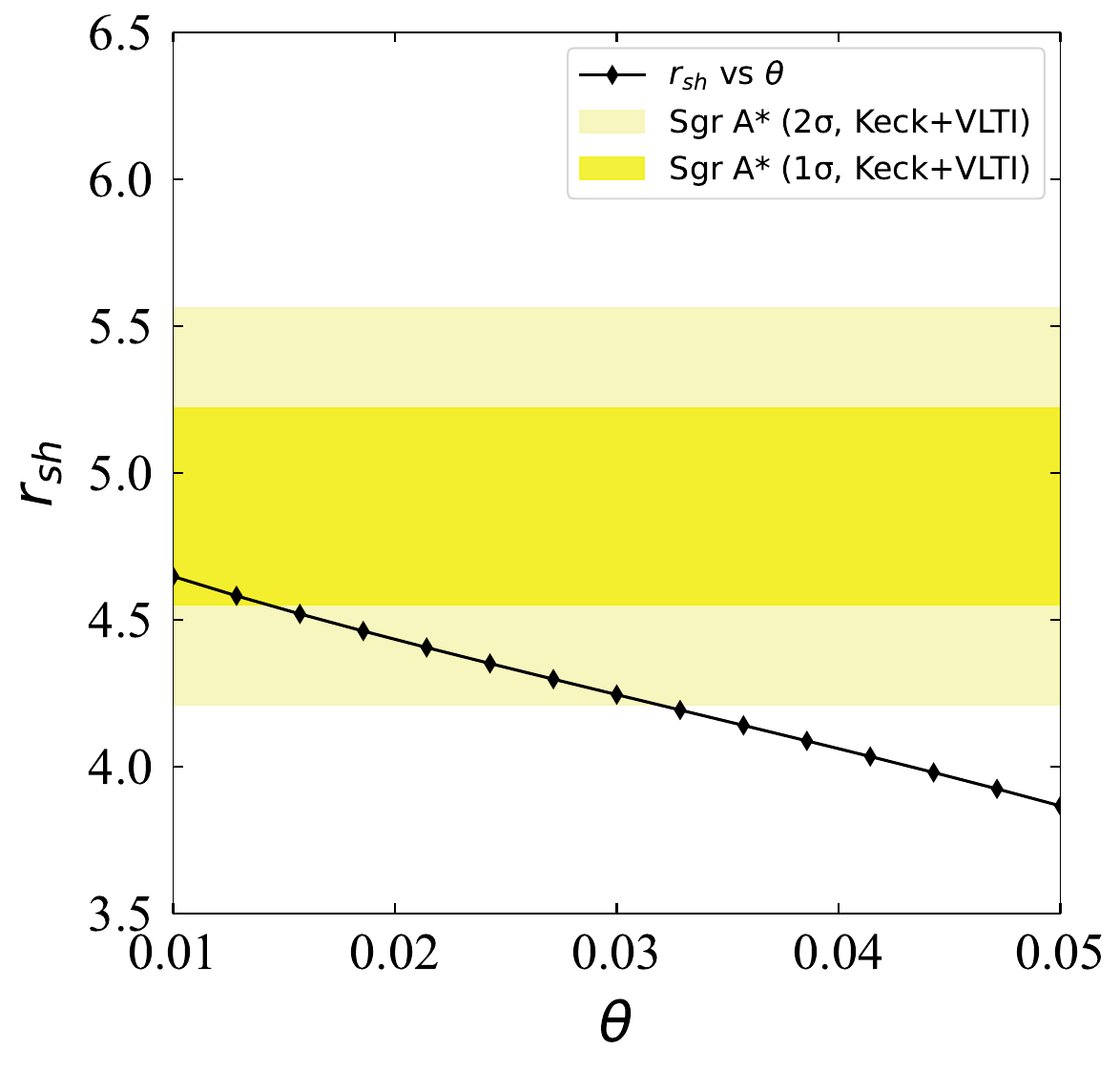}\hspace{0.1cm}\includegraphics[scale=0.3]{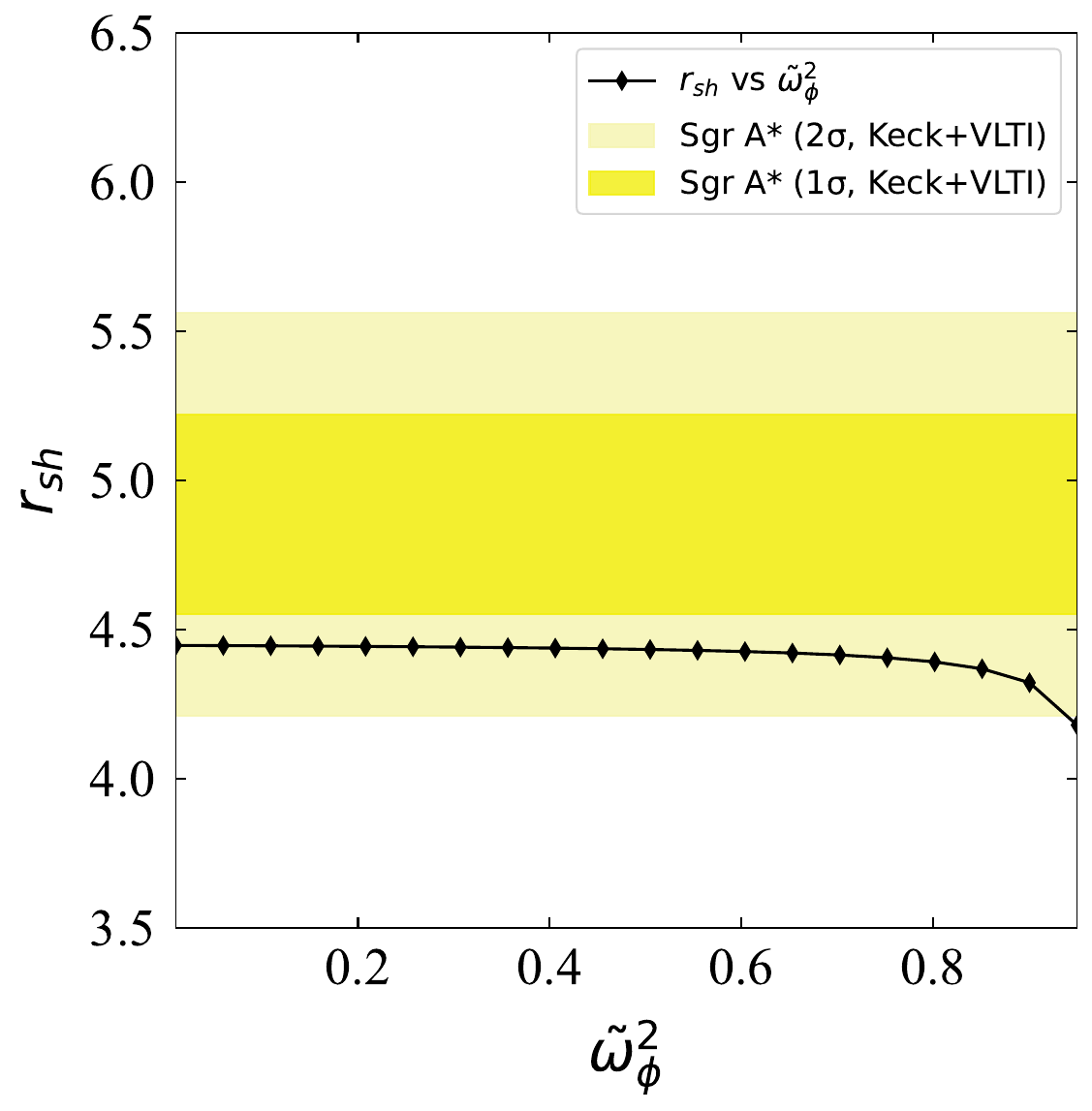} \hspace{0.1cm}\includegraphics[scale=0.3]{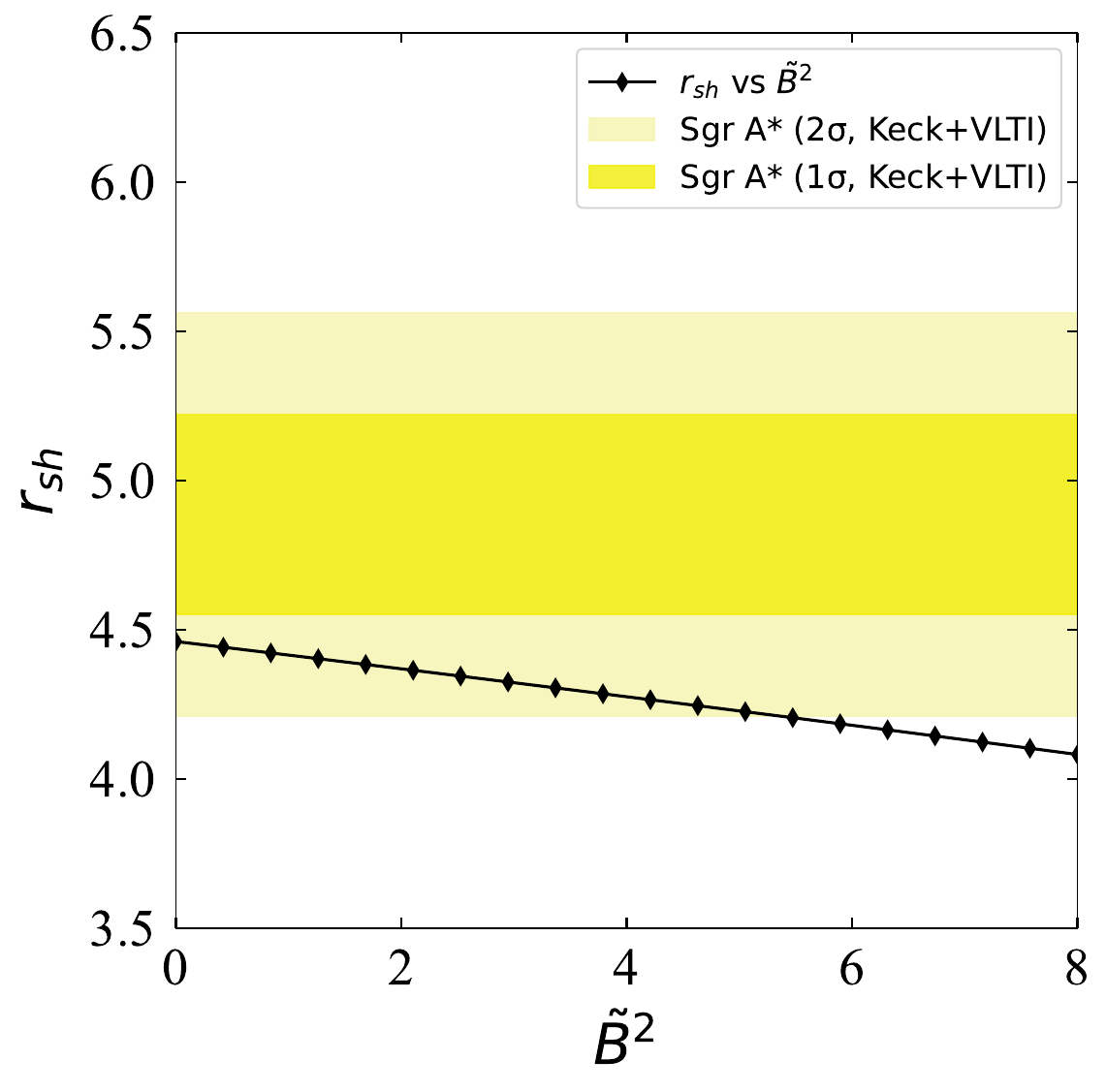}}
	\caption{Variation of shadow radius with the model parameters for inhomoegeneous plasma case with respect to the observational data. Here we have set $\omega_0^2 =1$, $\kappa^2 = 0.5$. Here we have set the observer at $r_o = 75$.}
\end{figure*}

To examine the shadow constraints on our model parameters, we use Eqs. \eqref{shad_rad_exp_homo} and \eqref{shad_rad_exp_inhomo} and plot it in comparison to the observational data as mentioned above. The Tab. \ref{tab1} shows the obtained constraints on the parameters with respect to the Sgr* shadow data. We note that for both homogeneous and inhomogeneous plasma cases, the non-commutative parameter $\theta$ is very tightly constrained on both 1$\sigma$ and 2$\sigma$ confidence levels. Moreover, the axion frequency $\tilde{\omega}_{\phi}$ is not constrained at 1$\sigma$ level for both types of plasma. Also, the magnetic field parameter $\tilde{B}^2$ remains unconstrained at the 1$\sigma$ confidence level for homogeneous plasma.

\section{Energy Emission from BH: Effect of Axion-Plasmon}
\label{sec6}
It is quite common in standard literature of particle scattering theory that the absorption cross section gives the probability of absorption. It is crucial to look into how different field or particles are absorbed and scattered close to BHs (see for instance Refs. \cite{Mashhoon1973May,Fabbri1975Aug,Page1976Jan,Unruh1976Dec,Jung2004Jul,
Doran2005Jun,Crispino2008Jul}). Moreover, the photon sphere can be interpreted as the capture cross-section of a BH and is a hypersurface made up of unstable null circular geodesics. Therefore, the BH shadow gives important information about the BH's high-energy absorption cross-section to an observer at infinity. This correlation results from the photon sphere's location at the maximum of the effective potential, which also happens to be the critical impact parameter for incoming null rays orbiting the photon sphere from infinity.

\begin{figure*}[htb]
\centerline{\includegraphics[scale=0.4]{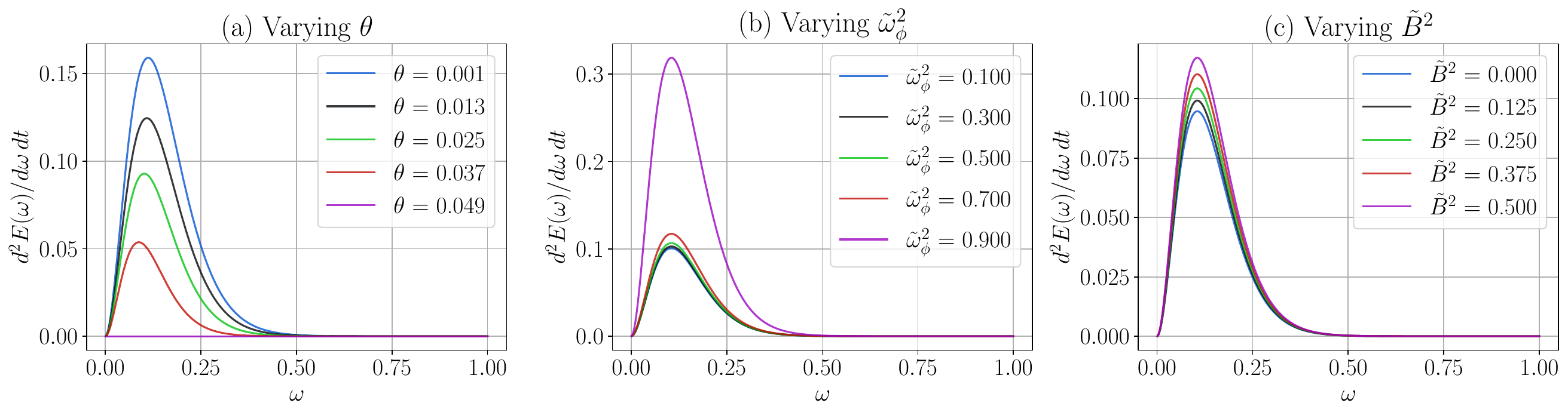}}
\caption{The energy emission by the BH is shown for the case of homogeneous plasma. Here we have fixed $\omega_0 = 1$, $\omega_p = 0.5$}
\label{ener_fig_homo}
\end{figure*}

\begin{figure*}[htb]
\centerline{\includegraphics[scale=0.4]{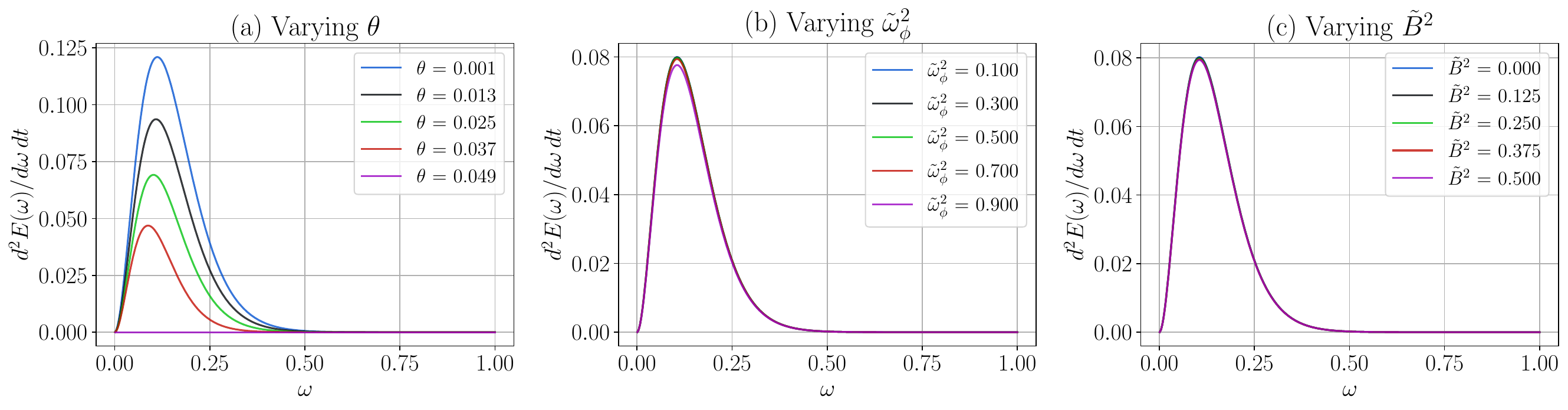}}
\caption{The energy emission by the BH is shown for the case of inhomogeneous plasma. Here we have fixed $\omega_0 = 1$, $\kappa = 0.5$}
\label{ener_fig_inhomo}
\end{figure*}

In this section, let us study the possible effect of the axion-plasmon on the energy emission rate from the BH. This can be directly estimated from the relation \cite{Khodadi2022Dec}
\begin{equation}
\frac{d^2 E (\omega)}{d\omega \, dt} = \frac{2\pi^2 \sigma_{lim}}{\exp(\omega/T)}\omega^3,
\label{energ}
\end{equation} 
where $T$ is the Hawking temperature, which is related to the surface gravity $\mathcal{K}$ through $ T = \frac{\mathcal{K}}{2\pi}$. In \cite{Sanchez1978Aug}, it was shown that the absorption cross-section oscillates around the limiting constant value, which is also directly related to the radius of the photon sphere and hence the shadow radius, i.e., the limiting constant $\sigma_{lim}$ is related to the shadow radius through
\begin{equation}
\sigma_{lim} \approx \pi r_{sh}^2.
\end{equation}
This reduces Eq. \eqref{energ} to:
\begin{equation}
\frac{d^2 E (\omega)}{d\omega \, dt} = \frac{2\pi^3 r_{sh}^2}{\exp(\omega/T) - 1} \omega^3.
\label{energ2}
\end{equation}
Having obtained Eq. \eqref{energ2}, one can numerically explore the behaviour of the energy emission with respect to the corresponding frequency $\omega$ of the photons (not to be confused with the plasma frequency). The Hawking temperature can be obtained from the usual relation $T = f'(r)/4\pi$, where $f(r)$ is given by Eq. \eqref{metric}. The shadow radii are imported from Eqs. \eqref{shad_rad_exp_homo} and \eqref{shad_rad_exp_inhomo} for homogeneous and inhomogenous plasma respectively. The numerical results are shown in the plots for both homogeneous and inhomogeneous plasma cases in Fig. \ref{ener_fig_homo} and \ref{ener_fig_inhomo} . 

From the results depicted in the plots we may see that in the homogeneous plasma case, the occurrence of peak height significantly depends on the non-commutative parameter $\theta$ (see Fig \ref{ener_fig_homo} (a) ). The maximum energy emission rate occurs when $\theta$ is lowest. With respect to varying $\tilde{\omega}_{\phi}^2$ in Fig \ref{ener_fig_homo} (b), the peak height seems to accelerate with the increasing values of $\tilde{\omega}_{\phi}^2$. Thus energy emission rate is the maximum for highest $\tilde{\omega}_{\phi}^2$. Lastly, in terms of varying $\tilde{B}^2$ as shown in Fig. \ref{ener_fig_homo} (c), similar behaviour is seen like the former case i.e., energy emission rate is maximum for highest value of $\tilde{B}^2$, however not very sensitive. 

Next for inhomogeneous plasma, a similar behaviour is inferred with respect to varying $\theta$, i.e, highest energy emission rate occurs for lowest $\theta$ (see Fig. \ref{ener_fig_inhomo} (a)). But in contrast to the homogeneous case, $\tilde{\omega}){\phi}^2$ and $\tilde{B}^2$ does not show observable effect on the energy emission rate, in other words the axion frequency and magnetic field has a minimal effect the energy emission rate (see Fig. \ref{ener_fig_inhomo} (b) and (c)).

\section{Conclusion}
\label{sec7}
To summarize, in this work  we have have systematically studied the influence of axion-plasmon on the photon sphere, BH shadows, and energy emission rates in the context of a non-commutative BH model. By taking into account both homogeneous and inhomogeneous plasma environments, we have investigated the effects of the associated model parameters, namely the non-commutative parameter \(\theta\), the axion frequency \(\tilde{\omega}_\phi^2\), and the magnetic field parameter \(\tilde{B}^2\).

For the investigation of the photon sphere radius, we found that in a homogeneous plasma, the radius increases exponentially with the plasma frequency, while larger values of \(\theta\) yield a decrease in the photon sphere radius. In contrast, although the axion and magnetic field parameters modify the rate of increase, their influence is however less pronounced at low plasma densities. In the inhomogeneous plasma scenario, where we assumed the plasma frequency following a power-law dependence, the radius initially increases with the coupling constant \(\kappa\) before approaching a saturation regime. The non-commutative parameter consistently exhibits a dominant effect, whereas the effects of \(\tilde{\omega}_\phi^2\) and \(\tilde{B}^2\) remain relatively subtle.

After studying the effect of the plasma on the photon sphere radius, we have used the results to study of BH shadows, including both angular and radial shadows. The angular shadow is found to decrease with the observer's distance and is notably sensitive to variations in \(\theta\). For homogeneous plasma, the axion frequency and magnetic field parameters modify the shadow behaviour predominantly at higher plasma frequencies, whereas in the inhomogeneous case, the observer's position is the primary dependent factor of the angular shadow.

To further back up our results, we compared our theoretical predictions with observational data from the Event Horizon Telescope, using which we have established constraints on the model parameters. In particular, the non-commutative parameter \(\theta\) is tightly constrained, whereas the axion and magnetic field parameters are less restrictive at the 1\(\sigma\) confidence level. It may be noted that Zeng et al \cite{Zeng2022Jan} studied the shadow properties of a non-commutative BH surrounded by accretion profiles, where they showed that non-commutative parameter plays an effective role in determining the shadow properties. In an another work, Campos et al \cite{Campos2022May} also studied shadows and quasinormal modes of a non-commutative BH obtained from a Lorentzian mass distribution. The magnitudes of the non-commutative parameter constrained in this study align with the aforementioned works. 

Finally, we studied the energy emission rate, which is done through the absorption cross-section associated with the photon sphere, shows that the peak emission is primarily influenced by the non-commutative parameter. In homogeneous plasma, both \(\tilde{\omega}_\phi^2\) and \(\tilde{B}^2\) contribute to an enhanced emission rate, while their effects are minimal in the inhomogeneous plasma case.

\bibliography{bibliography.bib}

\end{document}